\documentclass[aps,pra,floatfix,twocolumn]{revtex4}
\usepackage{amsmath,amssymb}
\usepackage{graphicx}
\usepackage{epsfig}
\usepackage{psfrag}
\usepackage[usenames]{color}
\usepackage{ulem}

\begin{document}

%\twocolumn
%\hsize\textwidth\columnwidth\hsize\csname@twocolumnfalse\endcsname

\title {Valley Hall transport of photon-dressed quasiparticles \\in two-dimensional Dirac semiconductors}

\author{V.~M.~Kovalev$^{1,2,3}$, Wang-Kong Tse$^{4}$, M.~V.~Fistul$^{3,5}$ and I.~G.~Savenko$^{3,6}$}
\affiliation {$^1$ A.V. Rzhanov Institute of Semiconductor Physics, Siberian Branch of Russian Academy of Sciences, Novosibirsk, 630090, Russia\\
$^2$ Department of Applied and Theoretical Physics, Novosibirsk State Technical University, Novosibirsk, 630073, Russia\\
$^{3}$ Center for Theoretical Physics of Complex Systems, Institute for Basic Science, Daejeon 34051, Republic of Korea \\
$^4$ Department of Physics and Astronomy, and Center for Materials for Information Technology, The University of Alabama, Alabama 35487, USA \\
$^{5}$ Russian Quantum Center, National University of Science and Technology "MISIS", 119049 Moscow, Russia\\
$^{6}$ Basic Science Program, Korea University of Science and Technology (UST), Daejeon 34113, Korea}
\date{\today}
\begin{abstract}
We present a theory of the photovoltaic valley-dependent Hall effect in a two-dimensional Dirac semiconductor subject to an intense near-resonant electromagnetic field. Our theory captures and elucidates the influence of both the field-induced resonant interband transitions and the nonequilibrium carrier kinetics on the resulting valley Hall transport in terms of photon-dressed quasiparticles. The non-perturbative renormalization effect of the pump field manifests itself in the dynamics of the photon-dressed quasiparticles, with a quasienergy spectrum characterized by {dynamical gaps $\delta_\eta$ ($\eta$ is the valley index)} that strongly depend on field amplitude and polarization. Nonequilibrium carrier distribution functions are determined by the pump field frequency $\omega$ as well as the ratio of intraband relaxation time $\tau$ and interband recombination time $\tau_{\mathrm{rec}}$. We obtain analytic results in three regimes, when (I) all relaxation processes are negligible, (II) $\tau \ll \tau_{\mathrm{rec}}$, and (III) $\tau \gg \tau_{\mathrm{rec}}$, and display corresponding asymptotic dependences on $\delta_\eta$ and $\omega$. We then apply our theory to two-dimensional transition-metal dichalcogenides, and find a strong enhancement of valley-dependent Hall conductivity as the pump field frequency approaches the transition energies between the pair of spin-resolved conduction and valence bands at the two valleys. 
\end{abstract}

\maketitle

%\newpage

%\subsection{Introduction}
\section{Introduction} 

Low-dimensional quantum systems subject to an externally applied large power high frequency electromagnetic field (EMF) display a great variety of interesting phenomena, such as multi-photon induced macroscopic quantum tunneling \cite{Ust1}, multi-photon Rabi oscillations and the dynamic Stark effect in superconducting or hybrid qubits \cite{Ust2,Bushev}, dissipationless electron transport~\cite{KibisPRL}, polaritons and condensates \cite{ExPol1,ExPol2}, and Floquet nonequilibrium states \cite{Flst1,Flst2}. In many cases of interest, the quantum dynamics of systems strongly interacting with an EMF can be described in terms of nonequilibrium quasiparticles called \textit{photon-dressed quasiparticles} (PDQs)~\cite{Ust2,MorinaPDQs}. They are characterized by a specific quasienergy spectrum and nonequilibrium steady-state distribution functions. Such a quasiparticle description is particularly useful for near-resonant excitation, i.e. when the frequency of the EMF is close to the difference of the intrinsic energy levels. The quasienergy spectrum of such PDQs shows a dynamical gap~\cite{ShelykhGap,Machlin} that is proportional to the amplitude of the EMF, and the nonequilibrium steady state of the PDQs is determined by interplay between different time scales: {the} inverse dynamical gap, inverse frequency, and relaxation times~\cite{BlochOsc}. 

As we turn to spatially extended systems, PDQs naturally appear in two-band semiconductors in the presence of EMF-induced interband  transitions. The quasiclassical dynamics of PDQs in a spatially dependent potential, for example, leads to a ballistic photocurrent in graphene-based nanostructures \cite{EfFist,SyzrFist,EfFistPol, RefKatsnelson}. 
The dependences of the photocurrent on the gate voltage, amplitude, frequency, and polarization of the EMF are mostly determined by the energy spectrum of the PDQs and, in particular, by the dynamical gap.  
However, a nonequilibrium steady state of PDQs cannot be achieved under these conditions, and thus has not been observed in such experiments.
Dynamical gaps have been extensively studied in originally gapless materials~\cite{RefSyzranov1, RefSyzranov2} under the high-power EMF, where rather complicated spectra of quasienergies with multiple dynamical gaps have been found.

In this paper we theoretically study the valley Hall transport of PDQs in homogeneous two-dimensional (2D)  Dirac semiconductors under irradiation of circularly polarized light, or in other words, a \textit{photovoltaic valley-dependent Hall effect}.  
%examine the influence of PDQs in 2D Dirac semiconductors by focusing on a specific transport phenomena, namely, the \textit{photovoltaic valley-dependent Hall effect}. 
It is well known that, in addition to momentum and spin, 2D materials with a hexagonal lattice (such as graphene \cite{Geim, RefKristinsson}) host valley degrees of freedom, which are quantum numbers describing corners $K$ and $K^{\prime}$ of their hexagonal Brillouin zone. {The presence of valleys gives rise to new valley-resolved physics \cite{xiao2007valley} that
%analogous to spin-resolved physics and 
has been much heralded as valleytronics~\cite{yao2008valley}.
%}
Two-dimensional Dirac semiconductors are gapped materials characterized by low-energy massive Dirac electrons in the vicinity of the two valleys. As an example of Dirac semiconductors, 2D transition-metal dichalcogenides (TMDs) \cite{jung2015origin,xu2014spin} provide a much sought-after platform to realize valley-resolved physics~\cite{RefMak, RefUbrig} due to a particularly large band gap, $\Delta$, advantageously occurring within the optical frequency range (e.g., MoS$_2$ has a band gap at $1.66\,\mathrm{eV}$ \cite{xiao2012coupled}). 
\begin{figure}[!t]
\includegraphics[width=\columnwidth]{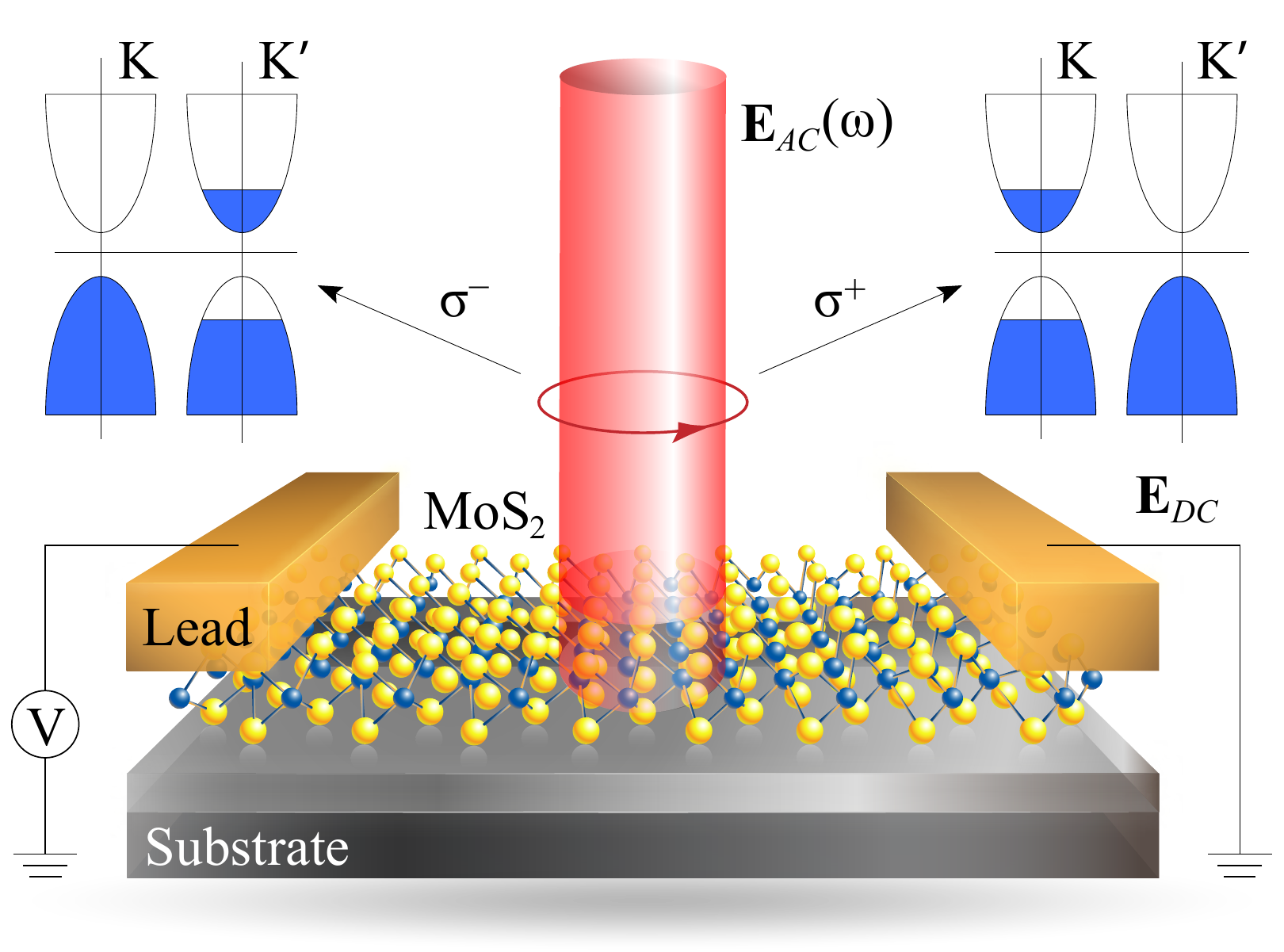} 
\caption{Schematic of EMF-induced Hall transport in a 2D Dirac semiconductor (monolayer MoS$_2$). $\textbf{E}_{DC}$ is a probe field (bias), and $\textbf{E}_{AC}$ is an external electromagnetic wave of light that can be right- or left-circular polarized. Depending on the polarization, either $K$ or $K^\prime$ valleys couple to light.  }
\label{Fig1}
\end{figure}

An important property underlying many valleytronic phenomena is the valley selection rule: the low-energy electrons at each valley couple predominantly to one particular state of optical 
polarization (left or right circular polarization), enabling valley-selective interband transitions.  
%This valley-selective optical transition rule is intimately 
%connected to the band topology of the massive Dirac Hamiltonian, which
%gives rise to opposite Berry curvatures at the two valleys. 
%These result 
%in a population imbalance of conduction band electrons between the two valleys, similar to a situation with majority and minority spins. 
Under a {DC} probe field, there will be an excess population of majority-valley electrons driven in the transverse direction, leading to an anomalous Hall effect. 
%This photovoltaic valley-dependent Hall effect is the subject of this paper. 
While the linear-response optical conductivity of TMDs has been extensively studied in a number of works (e.g., Refs.~\cite{gibertini2014spin,Li,Rostrami}), nonlinear optical phenomena \cite{NLBoyd} remain largely unexplored despite attracting increasing attention \cite{NLDirac1,NLDirac2}.

%In order to formulate a theory for the photovoltaic valley-dependent Hall conductivity, $\sigma_{xy}$, in this work we use a nonequilibrium Keldysh Green's function approach within the rotating wave approximation. Our theory leads to a generic expression for $\sigma_{xy}$ in 2D Dirac semiconductors that captures the dynamics of the PDQs, including their relaxation processes. We then consider three nonequilibrium regimes to investigate the importance of the intraband relaxation and interband recombination processes, and derive analytic results for valley-dependent photon-induced contributions to $\sigma_{xy}$. We numerically evaluate total $\sigma_{xy}$ for a specific 2D TMD case, and ultimately show that not only is $\sigma_{xy}$ greatly enhanced when the EF is near resonance, but also that the dependence of $\sigma_{xy}$ on EF amplitude and frequency can serve as a fingerprint for identifying different nonequilibrium steady states of PDQs. 

%-----------------------------------------------------------------------
%-----------------------------------------------------------------------
%-----------------------------------------------------------------------

\section{Results}

\subsection{Model, Hamiltonian and energy spectrum of photon-dressed quasiparticles} 

Let us consider the electron dynamics in a 2D Dirac semiconductor subjected to an externally applied strong pump EMF (Fig.~\ref{Fig1}), characterized by the vector potential $\textbf{A}(t)=\textbf{A}e^{-i\omega t}+\textbf{A}^*e^{i\omega t}$, where $\omega$ is the frequency of the applied field. The total Hamiltonian $\hat H$ of this system consists of two parts: the equilibrium Hamiltonian, 
\begin{equation}\label{eq1}
\hat H_0 =\frac{\Delta}{2}\hat {\sigma}_z +\hat {\textbf{v}}\cdot\hat {\textbf{p}}-\frac{1}{2}\lambda_\textrm{so} s\eta (\hat \sigma_z-1), 
\end{equation}
and the time-dependent Hamiltonian $\hat H_{\mathrm{int}} = ({e}/{c})\hat {\textbf{v}}\cdot\textbf{A}(t)$, describing the interaction of electrons with the EMF.
%and the time-dependent Hamiltonian, $\hat H_{int}$, describing the interaction of electrons with EF
%
%\begin{equation}\label{eq2}
%\hat H_{\mathrm{int}} =\frac{e}{c}\hat {\textbf{v}}\cdot\textbf{A}(t) . 
%\end{equation}
%
Here, $\hat {\textbf{p}}$ {and} $\hat {\textbf{v}}=(\hat{\text{v}}_x,\hat{\text{v}}_y)=\text{v}_0(\eta\hat\sigma_x,\hat\sigma_y)$ are the momentum and the single-particle velocity operator, respectively. The equilibrium Hamiltonian describes a pair of gapped Dirac cones (with the energy gap $\Delta$) at the two corners $K$ and $K^\prime$  of the hexagonal Brillouin zone (labeled by the valley index  $\eta= \pm 1$), and $\hat \sigma_{x,y,z}$ are Pauli matrices describing the pseudospin degrees of freedom. To apply our results to TMD materials, e.g. MoS$_2$,  we take into account the spin-orbit interaction $\lambda_{so}$ in the last term of the Hamiltonian in Eq.~(\ref{eq1}), with $s=\pm 1$ being the electron spin. Equation~\eqref{eq1} is the minimal model for TMD that captures valley Hall transport. Since the pump field is illuminated at near-resonant frequencies, effects from the conduction band edge spin splitting ($\sim~1$~meV) and trigonal warping further from the band edge ~\cite{RefDXiao,RefAsgari} are expected to be quantitatively small and can be neglected. 
% %
% \begin{figure}[!b] 
% \includegraphics[width=\columnwidth]{Fig2.pdf} 
% %[width=7cm, height=5cm]{Fig2.pdf} 
% \caption{Energy spectrum of electrons (in the absence of EF, $A=0$) and the quasienergy spectrum of photon-dressed quasiparticles. $K$ and $K^\prime$  are valleys of different spin, the hexagon in the center indicates the first Brillouin zone of the MoS$_2$. Red and blue dashed circles correspond to different polarizations of light. }
% \label{schematic2}
% \end{figure}
% %
In the absence of an external EMF, the electron energy spectrum of a TMD near $K$ and $K^\prime$ consists of conduction ($+$) and valence ($-$) bands that are spin and valley dependent, %with 
%$E_{s \eta }(p) = \sqrt{(\text{v}_0 p)^2+({\Delta-s\eta \lambda_{so}})^2/4}$. 
%four conduction bands $E_{s \eta }(p)$ and four valence bands $-E_{s \eta }(p)$ that are spin and valley dependent, with $E_{s \eta }(p) = \sqrt{(\text{v}_0 p)^2+({\Delta-s\eta %\lambda_{so}})^2/4}$. 
%
 \begin{equation}\label{eqspectrum}
 E_{s \eta }(p) =\pm \sqrt{(\text{v}_0 p)^2+
 \left (\frac{\Delta-s\eta \lambda_\textrm{so}}{2} \right )^2}.
 % E_{s \eta }(p) =\left [(\text{v}_0 p)^2+
 % \left (\frac{\Delta-s\eta \lambda_{so}}{2} \right )^2 \right ]^{1/2}.
 \end{equation}
%
%\textcolor{red}{It should be noted that Hamiltonian~\eqref{eq1} does not account for such effects as trigonal warping~\cite{RefMcCann, RefRozhkov} and additional spin %splitting~\cite{RefOrtiz, RefKurganova} like in graphene.}

In the presence of a pump EMF, it is convenient to introduce the quasienergy spectrum obtained by transforming the total Hamiltonian $\hat H$ to a rotating frame and neglecting all terms oscillating at the frequencies $2\omega$ within the rotating wave approximation (RWA). 
The RWA is valid for a near-resonant pump EMF, i.e. when $\omega \simeq 2|E_{s \eta}(p)|/\hbar$, and when the pump EMF amplitude is not too large, $e \text{v}_0 \vert \textbf{A} \vert/(\hbar \omega c) \ll 1$. This procedure has been previously used to obtain the quasienergy spectrum of weakly nonlinear oscillators \cite{Dykman}, electrons in two-band semiconductors \cite{EfFist,SyzrFist,TBSem,Oka}, and 2D electron gas with spin-orbit Rashba interaction \cite{FistEfetovSO}. 
%The RWA approach, however, neglects e.g. multiphoton processes in the presence of intense EF and as such RWA is a weak-coupling approximation \cite{RWA1,RWA2,Tse1}. The results obtained within RWA in this work are expected to hold as long  
%is consistent with exact results obtained from the Floquet approach in the weak  light-matter coupling regime. 
Thus, the quasienergy spectrum is given by (see Appendix~\ref{ApA}) 
\begin{gather}\label{eq5}
%\varepsilon_{1,2}(\textbf{p})=\frac{\lambda_{so}\eta s}{2}\pm\sqrt{\left(\Omega_{s\eta}(p)-\frac{\omega}{2}\right)^2+|\lambda_p|^2},
\varepsilon_{1,2}(\mathbf{p})=\pm\sqrt{\left(|E_{s \eta }(p)|-\frac{\hbar \omega}{2}\right)^2+\vert \delta_\eta(\mathbf{p})\vert^2}.
\end{gather}
Evidently, a strong pump EMF causes interband transitions with a { momentum- and valley-dependent} Rabi frequency, $\delta_\eta(\textbf{p})/\hbar$. The quasienergy spectrum in Eq.~(\ref{eq5}) for the PDQs is characterized by an opening of dynamical gaps, $\delta_\eta(\textbf{p})$,
%given by the Rabi frequency $\delta_R(\textbf{p})/\hbar$ 
as the resonant condition $|E_{s\eta}(p)|=\hbar \omega/2$ is satisfied. Dynamical gaps are generally anisotropic in the momentum space for elliptical EMF polarization, given by 
\begin{eqnarray}\label{eq6}
\delta_\eta(\mathbf{p}) &=&\frac{e\text{v}_0}{c}[\sin^2(\theta_p/2)e^{-i\eta\phi}(\eta A_x+iA_y) \\ \nonumber
&&+\cos^2(\theta_p/2)e^{i\eta\phi}(-\eta A_x+iA_y)],
\end{eqnarray}
where $\phi=\tan^{-1} (p_y/p_x)$ and $\sin \theta_p=\eta \text{v}_0 p/|E_{s \eta}(p)|$. We note that $\delta_\eta(\textbf{p})$ is proportional to {the amplitude $\vert \textbf{A}\vert$},
and it strongly depends on pump EMF polarization. In the vicinity of the valley centers, where $\text{v}_0p\ll{\Delta}$, one finds $|\delta_\eta(\textbf{p})| \simeq (e\text{v}_0/c)|-\eta A_x+iA_y|$. For valley Hall transport, we are interested in the circularly polarized pump field $\textbf{A}={A}_0(1,im)$, where $m=\pm 1$ is the helicity of the EMF. It follows from Eq.~(\ref{eq6}) that the magnitude of the dynamical gaps then becomes isotropic in the momentum space with $|\delta_\eta({p})|= (e\text{v}_0A_0/c)(1+\eta m \cos\theta_p)$, capturing the seminal valley-dependent selection rule~\cite{xiao2012coupled}. 
Therefore,  
%For simplicity, in the following we drop the complex modulus sign $\vert \dots \vert$ and denote $\delta_\eta({p})= (e\text{v}_0A_0/c)(1+\eta m \cos\theta_p)$, which is real and positive definite. 
while a dynamical gap opens in each of the four copies of the gapped Dirac dispersions in the TMD band structure, the valley selection rule causes a dynamical gap in one of the valleys to dominate. In what follows, we will write $\delta_\eta({p})$ instead of $|\delta_\eta({p})|$ thus dropping the irrelevant phase factor.
%
%\begin{figure}[!b] 
%\includegraphics%[width=7cm, height=7cm]%{Band_Dynamical_Gap_K_Kp_Quasibands.pdf} 
%[width=\columnwidth]{Band_Dynamical_Gap_K_Kp_Quasibands.pdf} 
%[width=7cm, height=5cm]{Fig2.pdf} 
%\caption{(a) Dynamical gaps in the rotating-frame conduction and valence bands under a left circularly polarized light. For clarity, here we only show the pair of energy dispersions for spin up electrons at valley $K$ and spin down electrons at valley $K^{\prime}$, which have a band gap $\Delta-\lambda_{\mathrm{so}}$. The other pair of energy dispersions with a band gap $\Delta+\lambda_{\mathrm{so}}$ will have dynamical gaps at a different value of momentum. (b) The corresponding quasienergy spectra  Eq.~(\ref{eq5}).}
%\label{DynGap}
%\end{figure}
%
%The magnitude of the dynamical gap then captures the seminal valley-dependent selection rule \cite{xiao2012coupled} and becomes isotropic in the momentum space, %$|\delta_R({p})|= (e\text{v}_0/c)(1+\eta m \cos\theta_p)$.  

%-----------------------------------------------------------------------
%-----------------------------------------------------------------------
%-----------------------------------------------------------------------

\subsection{Hall transport of photon-dressed quasiparticles}
%: Generic Results.} 
Hall transport in the presence of a strong pump EMF 
can be obtained as the linear response to a weak probe field of frequency $\Omega$ (see Fig.~\ref{Fig1}),  characterized by the vector potential 
$\mbox{\boldmath{$\mathcal{A}$}}(t)=\mbox{\boldmath{$\mathcal{A}$}}e^{-i\Omega t}$.  
The resulting current density is given by the expectation value $j_\alpha=-ie\,\textmd{Tr}[\hat{\text{v}}_{\alpha} G^<(t,t)]$, where $\alpha = x,y$ and $G^<$ is the lesser Green's function. In the linear regime over the probe field we obtain
\begin{gather}\label{eq3}
j_\alpha(t)=\int_{C} dt'Q_{\alpha\beta}(t,t')\mathcal{A}_{\beta}(t'),\\\nonumber
Q_{\alpha\beta}(t,t')=-i\frac{e^2}{c}\,\textmd{Tr}\left[\hat{\textrm{v}}_{\alpha} G(t,t')\hat{\textrm{v}}_{\beta} G(t',t)\right]_{C},
\end{gather}
where the times $t$, $t'$ are taken on the Keldysh contour $C$. The contour-ordered Green's functions $G(t,t')$ in~\eqref{eq3}, which are 2$\times$2 matrices due to the pseudospin structure of the Hamiltonian  (\ref{eq1}), are calculated by treating the pump field non-perturbatively within the RWA. 
\begin{figure}[b]
\includegraphics[width=7cm, height=4cm]{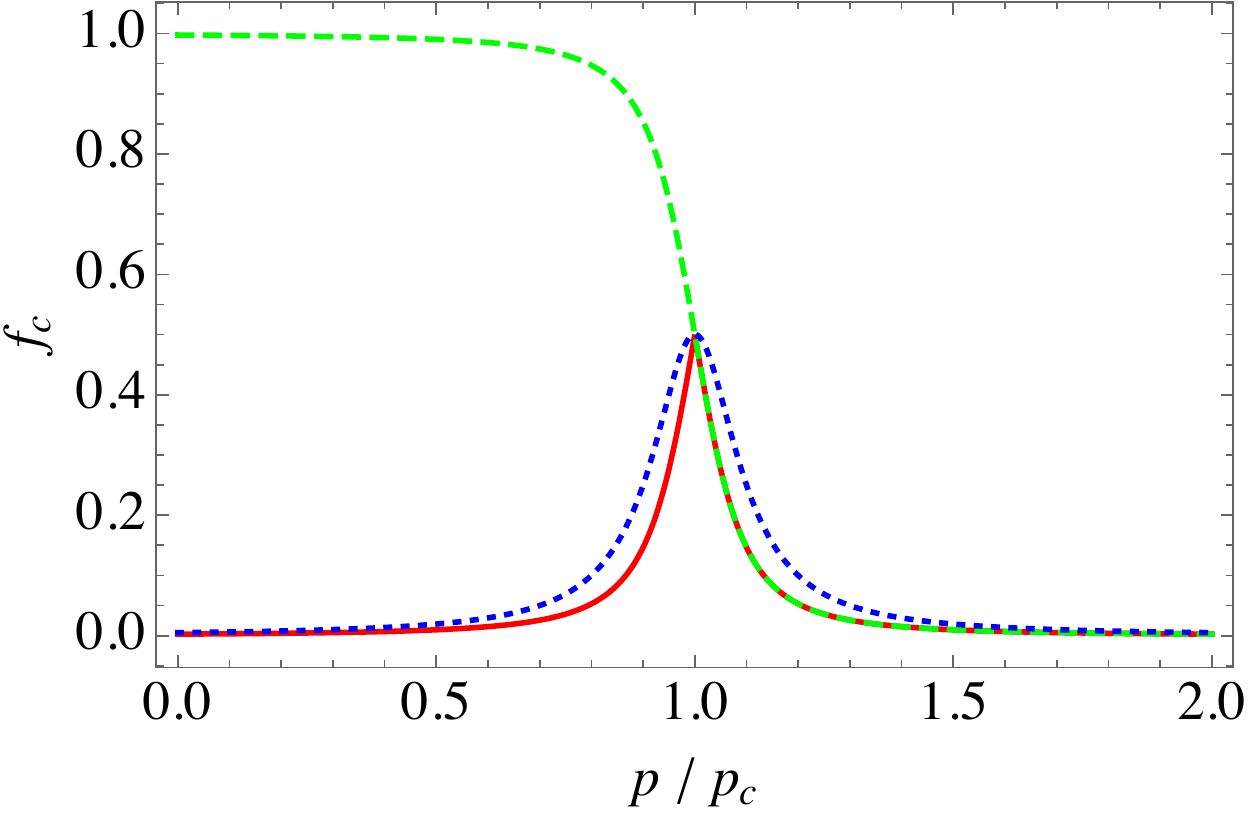} 
%[width=7cm, height=4cm]{Fig3.pdf} 
\caption{Nonequilibrium electron distributions $f_c$ as functions of the scaled momentum $p/p_c$, for $\hbar\omega =1.2 \Delta$, $\delta_\eta=0.2 \Delta$ and different values of relaxation and recombination times: regime I, $\tau^{-1}=\tau_{\mathrm{rec}}^{-1}=0$ (red); regime II, $\tau \ll \tau_{\mathrm{rec}}$ (green); and regime III, $\tau \gg \tau_{\mathrm{rec}}$ (blue). The momentum $p_c$ is the value at resonance and it is found from the relation: $\xi(p_c)=0$.}
\label{Fig2}
\end{figure}

The time-averaged Hall current is expressed via the Hall conductivity $\sigma_{xy}$ as $j_x=\sigma_{xy}\mathcal{E}_y$, {and $\mathcal{E}_y(t)=-\frac{1}{c}\partial_t\mathcal{A}_y(t)$ } {is the probe electric field} taken along the $y$ axis. 
%The Hall conductivity  $\sigma_{xy}$ contains the nonlinear effects due to the presence of a strong pump EF.  
Following calculation given in Appendix~\ref{ApA}, %(details are presented in the Appendix), 
we find in the limit of a static probe field ($\Omega \rightarrow 0$, thus $\mathcal{E}_y\equiv E_{DC}$) a generic expression for the \textit{photovoltaic valley-dependent Hall conductivity}: 
\begin{gather}\label{eqSuper}
\sigma_{xy} = \frac{2e^2\text{v}_0^2}{\hbar}\sum_{s,\eta} \eta \int\frac{d^2\textbf{p}}{(2\pi)^2}\cos\theta_\textbf{p} [n_2(\textbf{p})-n_1(\textbf{p})]\\\nonumber
\times\left[\frac{u^4_{\textbf{p}}}{(\varepsilon_1(\textbf{p})-\varepsilon_2(\textbf{p})+\hbar\omega)^2}-\frac{v^4_{\textbf{p}}}
{(\varepsilon_1(\textbf{p})-\varepsilon_2(\textbf{p})-\hbar\omega)^2}\right],
\end{gather}
where the coefficients $u_\textbf{p}$ and $v_\textbf{p}$ satisfy the following conditions (we will use $p$ instead of $\textbf{p}$ in the indices in what follows): $u^2_{p}+v^2_{p}=1$, $u^2_p-v^2_p=\sqrt{\xi^2(p)+\delta^2_\eta(p)}/\xi$,  $\xi(p)=|E_{s\eta}(p)|-\hbar\omega/2$, and $n_{1,2}({p})$, which are the nonequilibrium distribution functions of the PDQs. The nonequilibrium electronic distribution functions of the conduction and valence bands,  $f_c(p)$ and $f_v(p)$ (see Fig.~\ref{Fig2}), are related to those of the PDQs as 
\begin{gather}\label{EqDF}
f_c({p})=u_p^2n_1({p})+v_p^2n_2({p}),\\\nonumber
f_v({p})=u_p^2n_2({p})+v_p^2n_1({p}).
\end{gather}
Since $f_c({p})+f_v({p})=1 $ by particle number conservation, the above equation implies the conservation of PDQs with $n_1(p)+n_2(p) = 1$. The valley-dependent Hall conductivity in Eq.~(\ref{eqSuper}) depends on the population difference of the PDQs, which is given by $n_2({p})-n_1({p}) =  1-2n_1({p})=[1-2f_c({p})]{\sqrt{\xi^2(p)+\delta_\eta^2(p)}}/{\xi(p)}$. 

Equation~\eqref{eqSuper} contains two contributions that are due to the resonant and nonresonant interaction of electrons with the EMF. The resonant contribution to the Hall conductivity is determined by a narrow region of $\xi \simeq \delta_\eta (p_{c})$, where the RWA is well justified. Here, $p_{c}$ is the solution of equation $\xi(p_{c})=0$. 

The nonresonant contribution to the valley Hall conductivity stems from a broad region of $\xi \simeq \hbar\omega$ in Eq.~\eqref{eqSuper}, and for small values of $\delta_\eta \ll \hbar\omega$ the nonresonant interaction leads just to small corrections to the dark value of $\sigma^{eq}_{xy}=e^2/(4\pi\hbar)$ calculated in the absence of EMF~\cite{Asgari}. These small corrections cannot be elaborated precisely in the  framework of RWA but their typical value $\propto ( \delta_\eta)^2/(\hbar\omega)^2$ is smaller than the resonant contribution to the Hall conductivity (see Sec. D). Thus one can safely omit the influence of the nonresonant interactions on the valley Hall conductivity in 2D wide gap Dirac semiconductors. 

%While the photovoltaic Hall effect has been recently studied in a gapless Dirac system (\textit{i.e.}, graphene) illuminated by an intense circularly polarized light \cite{Oka}, our result Eq.~(\ref{eqSuper}) applies in the complementary case of a gapped Dirac system described by the massive Dirac Hamiltonian.  
% It should be mentioned that the photovoltaic Hall effect has recently been studied in gapless graphene illuminated by an intense circularly polarized light using Floquet theory \cite{Oka}. Our result Eq.~(\ref{eq4}) applies for gapped Dirac systems described by the massive Dirac Hamiltonian.  
%By focusing on the near-resonant and weak coupling regime, our RWA approach allows to obtain analytic results often unavailable from formally and numerically exact Floquet analysis. Therefore, it provides important complementary physical insight as will be discussed in the following sections. 

%----------------------------------------
%----------------------------------------
%----------------------------------------

\subsection{Kinetics of photon-dressed quasiparticles}
We consider an insulating Dirac semiconductor in equilibrium, where the Fermi level is located in the middle of the band gap. The temperature is taken to be much smaller than the band gap so that thermally excited carriers can be ignored. In the presence of a strong pump EMF, the nonequilibrium distribution function of electrons depends on the ratio of the intraband relaxation time $\tau$ and the interband recombination time $\tau_{\mathrm{rec}}$ \cite{Galitskii,elesin1971coherent,Remark_Elesin}. In the absence of any intraband relaxation and interband recombination, i.e. the ballistic regime (later referred to as regime~I), the difference in the distribution functions of the PDQs is given by $1-2n^{(\mathrm{I})}_1({p})=\textmd{sign}(\xi)$, corresponding to the distribution function of nonequilibrium electrons in the conduction band, $f^{(\mathrm{I})}_c({p})=({1}/{2})\left[1-{|\xi(p)|}/{\sqrt{\xi^2(p)+\delta_\eta^2(p)}}\right]$.  
Here, our results coincide with a kinetic equation analysis based on the density matrix approach~\cite{Haug,Tse}. 

Under a strong pump field with large Rabi frequency  $\delta_\eta/\hbar \gg \max\{1/\tau, 1/\tau_{\mathrm{rec}}\}$, various nonequilibrium distributions of PDQs can be achieved. 
%{As the pump EF is rather large, and more precisely $\delta_\eta \gg 
%\max\{\hbar/\tau, \hbar/\tau_{\mathrm{rec}}\}$, the various non-equilibrium distributions of PDQs can be achieved. } 
If the intraband scattering time is small such that $\tau \ll \tau_{\mathrm{rec}}$ (regime II or inverted population regime), we have $n^{(\mathrm{II})}_1({p})=0$ and $1-2n^{(\mathrm{II})}_1({p})=1$, corresponding to $f^{(\mathrm{II})}_c({p})=v_p^2=({1}/{2})\left[1-{\xi(p)}/{\sqrt{\xi^2(p)+\delta_\eta^2(p)}}\right]$. In the opposite case, when the interband recombination time is small, $\tau \gg \tau_{\mathrm{rec}}$ (regime III), 
we have $n^{(\mathrm{III})}_1({p})=v_p^2$ and $1-2n^{(\mathrm{III})}_1({p})=u_p^2-v_p^2={\xi(p)}/{\sqrt{\xi^2(p)+\delta_\eta^2(p)}}$, corresponding to  $f^{(\mathrm{III})}_c({p})=2u_p^2v_p^2={\delta_\eta^2(p)}/{2\left[\xi^2(p)+\delta_\eta^2(p)\right]}$.  Note here that the nonequilibrium state of the PDQs in regime III is analogous to a nonequilibrium steady state of a two-level system subject to a strong resonant EMF~\cite{Blochequation}.

%-----------------------------------------------------------------------
%-----------------------------------------------------------------------
%-----------------------------------------------------------------------

\subsection{Nonequilibrium valley-resolved Hall conductivity} 
In order to focus on the essential valley-resolved physics, we will first disregard the spin-orbit interaction.
%in the present section. 
%For simplicity of discussion we neglect spins and the spin-orbit interaction; 
%If they are included, one can simply substitute $\Delta \to \Delta-s\eta \lambda_{\mathrm{so}}$ and then sum over the resulting Hall conductivity over the spin degrees of freedom. We will restore the spin-orbit interaction when we evaluate the full $\sigma_{xy}$ numerically in the section thereafter.  
%replace the following analytic results simply by the substitution $\Delta \to \Delta-s\eta \lambda_{\mathrm{so}}$ and sum the Hall conductivity over the spin degrees of freedom. 
%For definiteness, we take the pump EF to be left circularly polarized ($m = 1$).  The valley selection rule then implies that the pump field will couple only to the $K$ valley ($\eta=1$) while the $K^\prime$ 
%%valley ($\eta=-1$) remains approximately uncoupled, so that the Hall conductivity at the $K^\prime$ 
%$K^\prime$ 
%valley can be treated as $\sigma_{xy}^{s,{K}^\prime} \approx \sigma_{xy}^{s,{K}^\prime}(\delta_R = 0)$.
Then the Hall conductivity in valley $\eta$ reads
\begin{equation}\label{eq8}
\sigma_{\eta, xy}=\frac{\eta e^2{\Delta}}{32\pi \hbar}\int\limits_{-({\hbar\omega-{\Delta}})/{2}}^\infty
d\xi[n_2({p})-n_1({p})]\mathcal{F}_\omega(\delta_\eta,\xi),\\ 
\end{equation}
where
\begin{equation}\label{eq81}
\mathcal{F}_\omega(\delta_\eta,\xi)=
\left(\frac{1+\frac{\xi}{\sqrt{\xi^2+\delta_\eta^2}}}{\sqrt{\xi^2+\delta_\eta^2}+\frac{\hbar\omega}{2}}\right)^2-
\left(\frac{1-\frac{\xi}{\sqrt{\xi^2+\delta_\eta^2}}}{\sqrt{\xi^2+\delta_\eta^2}-\frac{\hbar\omega}{2}}\right)^2. 
\end{equation}
%
%Thus, one can see that the Hall conductivity is determined by non-equilibrium distribution functions of PDQs{, $n_{1,2}({p})$}. 
In the limit of vanishing pump EMF, i.e. $\delta_\eta \to 0$ (and for arbitrary frequency), the distribution functions  reduce to those in equilibrium, $f_c({p})=0$ and $1-2n_1({p})=1$, 
so that Eq.~(\ref{eq8}) recovers the correct value of the dark Hall conductivity of a single valley~\cite{Asgari}, $\sigma^{eq}_{xy}= \eta e^2/(4\pi \hbar)$. 
Substituting the expressions for $n_{1,2}(p)$ in Eq.~(\ref{eq8}), we obtain 
%For these three regimes the $\eta$-valley contribution to the  Hall conductivity is written as 
%contribution Taking the sum over the two valleys, we find the following total valley-dependent Hall conductivity in the three regimes
%
\begin{gather}
\label{eqMain}
\left(
                   \begin{array}{c}
                   \sigma_{\eta,xy}^{(\mathrm{I})}\\
                    \sigma_{\eta,xy}^{(\mathrm{II})} \\
                    \sigma_{\eta, xy}^{(\mathrm{III})} \\
                   \end{array}
                 \right)
=\frac{\eta e^2{\Delta}}{32\pi \hbar}\int\limits_{-(\hbar \omega-{\Delta})/2}^\infty
d\xi
\left(
  \begin{array}{c}
  \textmd{sign}\xi\\
    1 \\
    \frac{\xi}{\sqrt{\xi^2+\delta_\eta^2}} \\
  \end{array}
\right)\mathcal{F}_\omega(\delta_\eta,\xi). 
\end{gather}
%
% %
% \begin{gather}
% \label{eqMain}
% \left(
%                    \begin{array}{c}
%                    \sigma_{\eta,~ xy}^{(\mathrm{I})}\\
%                     \sigma_{\eta,~ xy}^{(\mathrm{II})} \\
%                     \sigma_{\eta,~ xy}^{(\mathrm{III})} \\
%                    \end{array}
%                  \right)
% =\eta \frac{e^2{\Delta}}{32\pi \hbar}\int\limits_{-(\omega-{\Delta})/2}^\infty
% d\xi\\\nonumber
% \times
% \left[\left(
%   \begin{array}{c}
%   \textmd{sign}\xi\\
%     1 \\
%     \frac{\xi}{\sqrt{\xi^2+\delta_\eta^2}} \\
%   \end{array}
% \right)\mathcal{F}_\omega(\delta_\eta,\xi)-(\textmd{sign}\xi)\mathcal{F}_\omega(0,\xi)\right]+\frac{\eta e^2}{4\pi \hbar}. 
% \end{gather}
% %

Further analytic progress can be made if we disregard the dependence $\delta_\eta$ on $p$. %and focus on the frequency range $\vert \hbar \omega-\Delta \vert \gg \delta_\eta$. 
Indeed, $\delta_\eta (p)$ is a smooth function. In the mean time, the main contribution to the nonequilibrium part of the valley Hall conductivity in Eq.~\eqref{eqMain} comes from the vicinity of the resonant points. Thus we substitute the dependence $\delta_\eta (p)$ by the value $\delta_\eta=\delta_\eta (p_{c})$. (It should be noted that we keep the dependence of $\delta_\eta$ on frequency $\omega$.)

%Using Eq. (8) for the K-valley which is strongly coupled to the left circularly polarized light , and completely neglecting the interaction with the light in the $K^\prime$ -valley, we numerically calculateâŠ Next we take into account the light-matter coupling at both valleys $K$ and $K^\prime$...

\begin{widetext}
\begin{table}[t!]
\centering
\begin{tabular}{|c|c|c|}
\hline
Regime I                                                                                                                 & Regime II                                                                                                                  & Regime III                                                                                                                                                 \\ \hline
$\tau,~\tau_{\textrm{rec}}=\infty$                                                                                       & $\tau\ll\tau_{\textrm{rec}}$                                                                                               & $\tau\gg\tau_{\textrm{rec}}$                                                                                                                               \\ \hline
\begin{tabular}[c]{@{}c@{}}Dynamical gap opens (for 
one valley):
\\ Thresholdlike behavior at $\hbar\omega=\Delta$\end{tabular}  & \begin{tabular}[c]{@{}c@{}}Dynamical gap opens (for 
one valley): \\ Giant increase of $\sigma_{xy}$ after $\hbar\omega=\Delta$\end{tabular} & \begin{tabular}[c]{@{}c@{}}Dynamical gap opens (for 
one valley): \\ Thresholdlike $\tau_{\textrm{rec}}$-dependent behavior \\ at $\hbar\omega=\Delta$\end{tabular} \\ \hline
\begin{tabular}[c]{@{}c@{}}Dynamical gaps open (for both valleys): \\ Thresholdlike behavior at $\hbar\omega=\Delta$\end{tabular} & \begin{tabular}[c]{@{}c@{}}
Dynamical gaps open (for both valleys): \\ Strong compensation of $\sigma_{xy}$'s from both valleys \end{tabular}             & \begin{tabular}[c]{@{}c@{}}Dynamical gaps open (for both valleys): \\ Thresholdlike behavior at $\hbar\omega=\Delta$\end{tabular}                                   \\ \hline
\end{tabular}
\caption{Summary of the three regimes.}
\label{my-label}
\end{table}
\end{widetext}

Furthermore let us focus on the frequency range $\vert \hbar \omega-\Delta \vert \gg \delta_\eta$.
If the pump EMF frequency is below the gap, ${\Delta}-\hbar \omega \gg \delta_\eta$, only virtual transitions between the conduction and valence bands 
%in the K valley 
occur, resulting in a renormalization of band energies, i.e. the dynamic Stark effect, as described by the quasienergies  $\varepsilon_{1,2}(p)$ of the PDQs. This scenario corresponds to regime I. 
%and it results just in the renormalization of  PDQs energies $\epsilon_{1,2}(p)$,  i.e. the  \textit{dynamic Stark effect}. However, in this frequency range the regime I  can be %realized only. 
Calculating the integral over $\xi$ in Eq.~(\ref{eqMain}), we obtain $\sigma_{\eta,xy}=\sigma^{eq}_{\eta,xy}+\sigma^{neq}_{\eta,xy}$, and the $\eta$-valley nonequilibrium contribution to the Hall conductivity $\sigma^{neq}_{\eta,xy}$ as 
\begin{gather}\label{eq11}
\sigma_{\eta,xy}^{neq,(\mathrm{I})}=-\frac{\eta e^2}{2\pi \hbar}\frac{\delta^2_{\eta}}{{\Delta}({\Delta}-\hbar\omega)}.
\end{gather}
%
%
% \begin{gather}\label{eq11}
% \sigma_{\eta,~xy}^{(\mathrm{I})}=-\eta\frac{e^2}{2\pi \hbar}\frac{\delta^2_R}{{\Delta}({\Delta}-\omega)}+\frac{\eta e^2}{4\pi \hbar}.
% %\sigma_{xy}^{(\mathrm{I})}=-\frac{e^2}{4\pi}\frac{2\delta^2_R}{{\Delta}({\Delta}-\omega)},~~{\Delta}-\omega \gg 2\delta_R
% \end{gather}
%

In the opposite limit ($\omega>\Delta/\hbar$), interband transitions occur and all three regimes can be established. 
\begin{figure}[!b]
\includegraphics[width=7cm, height=4cm]{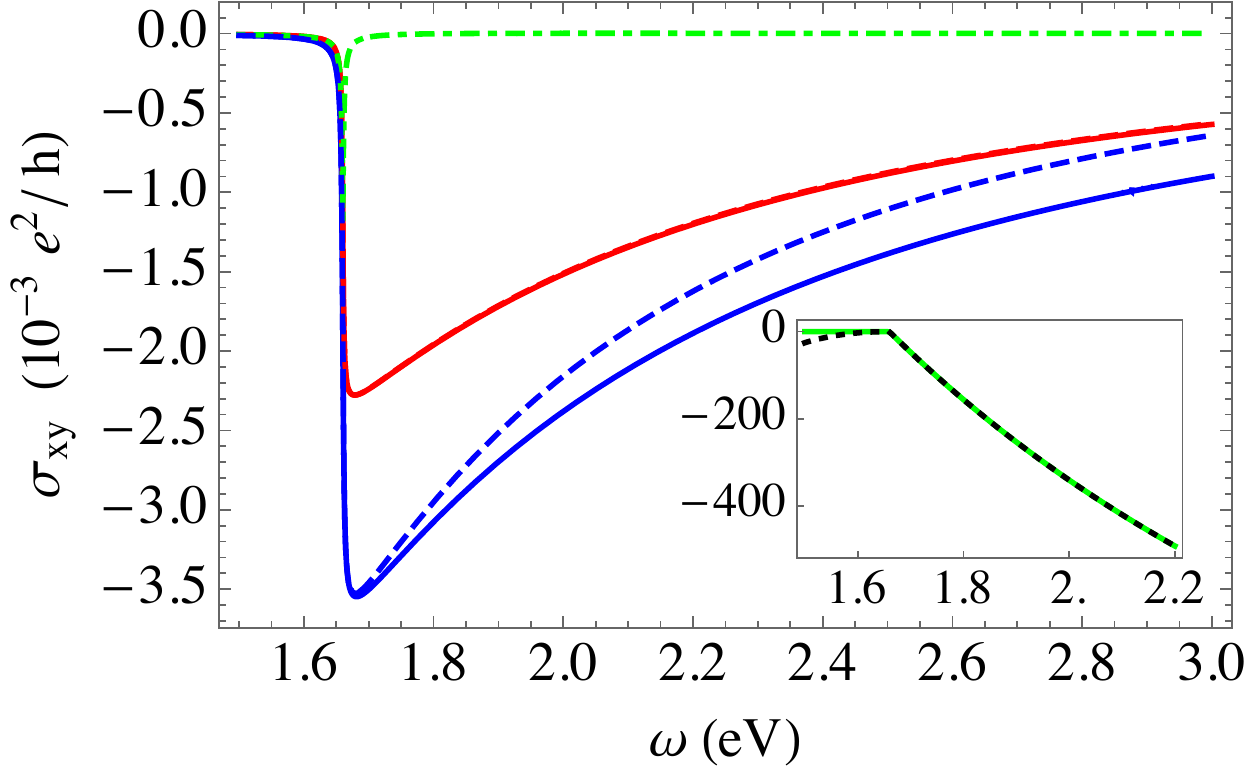}
%[width=7cm, height=4cm]{Fig4_new.pdf}
\caption{Photovoltaic Hall conductivity as a function of pump frequency in Regimes I (red), II (green), and III (blue). Calculations were performed using Eq.~(\ref{eq8}) (solid lines) and Eq.~(\ref{eqSuper}) (dashed lines), with the latter case taking into account the full momentum dependence of $\delta_{\eta}(p)$. The dynamical gap $\delta_{1} (p=0) = 0.5$ meV was chosen. Red dashed and solid curves coincide. Inset shows $\sigma^\textrm{(II)}_{1,xy}$ calculated using Eq.~\eqref{eq8} accounting for nonequilibrium particle distribution only in single valley. Black dashed curve in Inset is analytic result via Eq.~\eqref{eq13_2} (applicable if $\hbar\omega>\Delta$).}
 \label{Fig3}
\end{figure}
Calculating the integral over $\xi$ in Eq.~(\ref{eqMain}), we arrive at the following results: 
\begin{gather}
\sigma^{neq,(\mathrm{I})}_{\eta,xy}(\omega)= -\frac{\eta e^2}{\pi \hbar}\frac{\Delta \delta_\eta}{(\hbar\omega)^2} ,\label{eq13_1} \\
%-\frac{\eta e^2}{\pi \hbar} \frac{\Delta \delta_\eta}{\omega^2}+\frac{\eta e^2}{4\pi \hbar} ,\label{eq13_1} \\
\sigma^{neq,(\mathrm{II})}_{\eta, xy}(\omega)=\frac{\eta e^2}{2\pi \hbar}\left[-1+\frac{{\Delta}}{\hbar\omega}-\frac{\delta_\eta^2}{\Delta(\hbar \omega-\Delta)} \right]\label{eq13_2},\\
%\frac{\eta e^2}{2\pi \hbar}\left[\frac{{\Delta}}{\omega}-\frac{\delta_\eta^2}{\Delta(\omega-\Delta)} \right]\label{eq13_2},\\
\sigma^{neq,(\mathrm{III})}_{\eta,xy}(\omega)=-\frac{\eta e^2}{2\pi \hbar}\frac{\pi \Delta\delta_\eta}{(\hbar\omega)^2}. \label{eq13_3}
\end{gather}
%
%
% \begin{gather}
% \sigma^{(\mathrm{I})}_{\eta,~xy}(\omega)=-\frac{\eta e^2}{\pi \hbar} \frac{\Delta \delta_\eta}{\omega^2}+\frac{\eta e^2}{4\pi \hbar} ,\label{eq13_1} \\
% \sigma^{(\mathrm{II})}_{\eta, ~xy}(\omega)=-\frac{\eta e^2}{4\pi \hbar}\left[\frac{\omega-2{\Delta}}{\omega}+\frac{2\delta_\eta^2}{\Delta(\omega-\Delta)} \right]+\frac{\eta e^2}{4\pi \hbar}\label{eq13_2},\\
% \sigma^{(\mathrm{III})}_{\eta,~xy}(\omega)=-\frac{\eta e^2}{2 \hbar}\frac{\Delta\delta_\eta}{\omega^2}+\frac{\eta e^2}{4\pi \hbar} . \label{eq13_3}
% \end{gather}
%

%
%
%
\begin{figure*}[!t]
\includegraphics[width=0.31\linewidth]{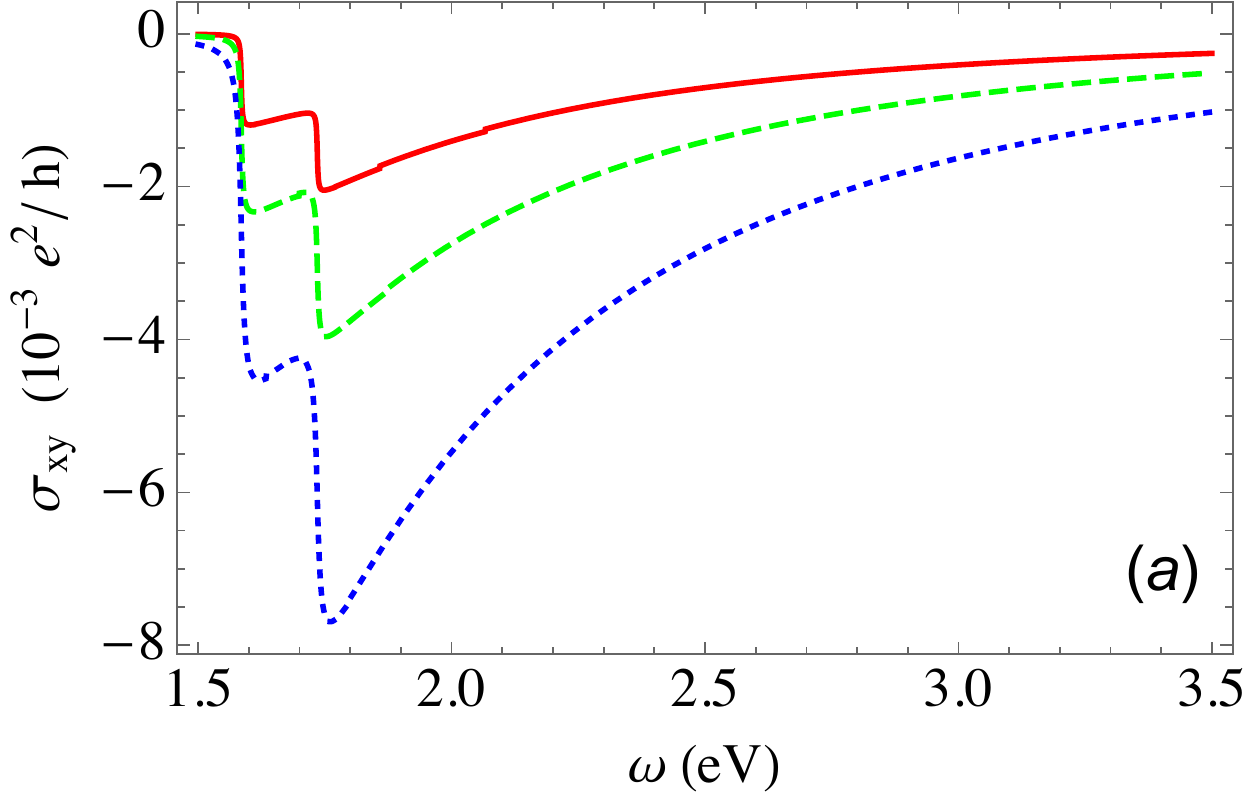}
\includegraphics[width=0.32\linewidth]{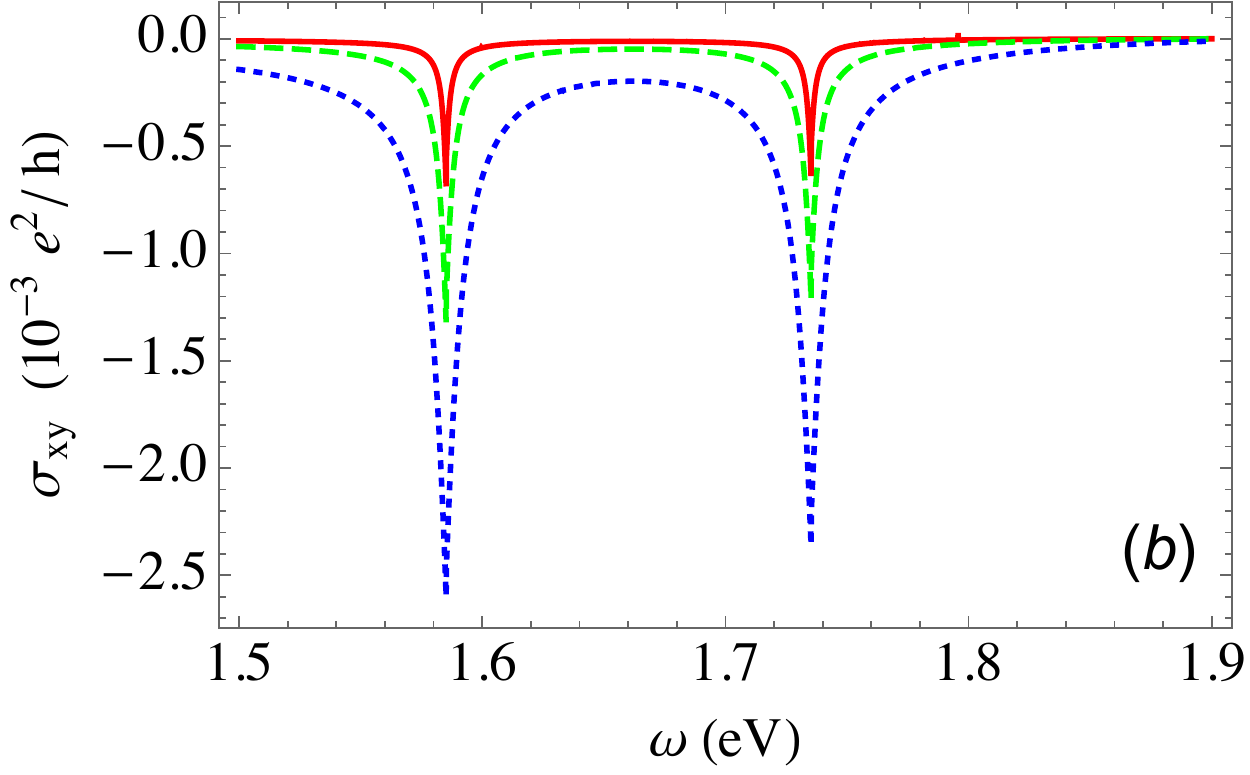}
\includegraphics[width=0.32\linewidth]{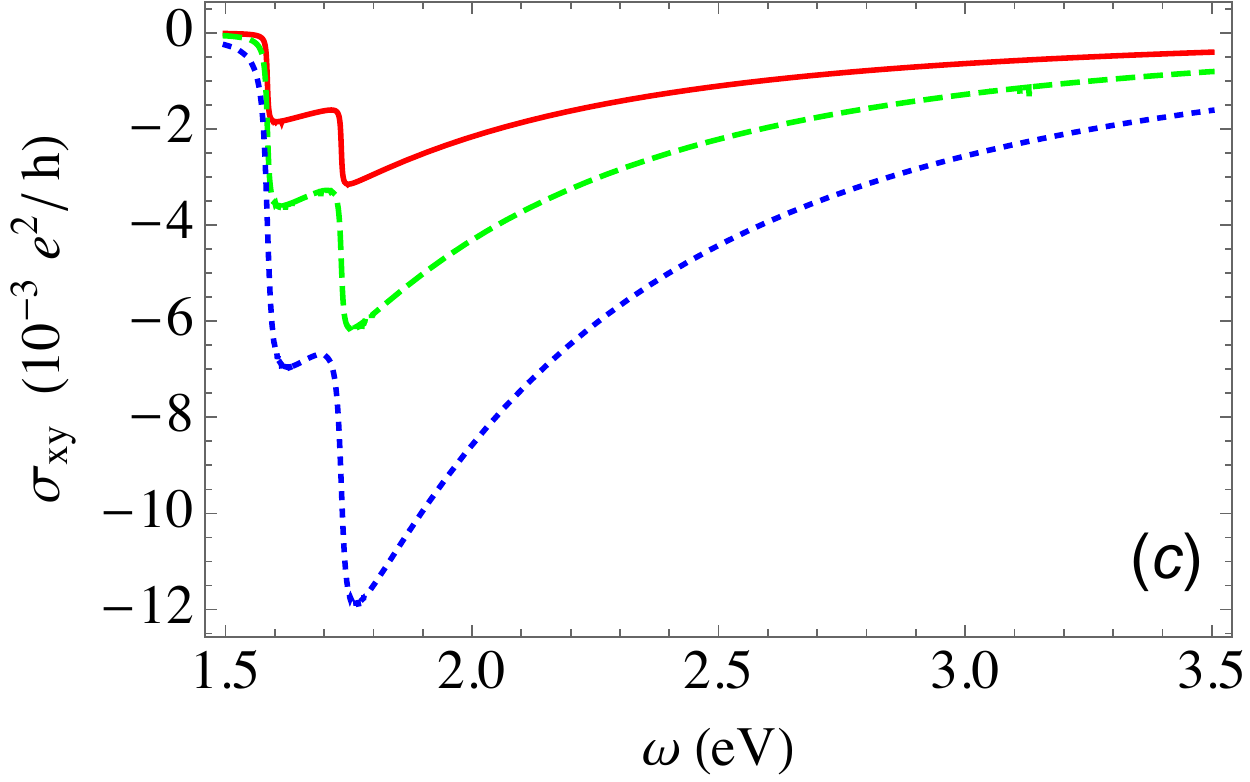}
\caption{Photovoltaic Hall conductivity $\sigma_{xy}(\omega)$ for (a) regime I, (b) regime II, and (c) regime III. 
Red solid, green dashed, and blue dotted curves correspond to $\delta_{\eta}(0)=0.5$, $1$, and $2$ meV, respectively.} 
\label{Fig4}
\end{figure*}
We see opposite signs of the nonequilibrium $\sigma^{neq}_{xy}$ and dark $\sigma^{eq}_{xy}$ contributions to the Hall conductivity---this is intimately connected with band topology and the sign of the Berry curvatures. Indeed,  
%Under a left circularly polarized pump field ($m = 1$), the total Hall conductivity is negative in all three regimes. 
%the negative sign of the total Hall conductivity in all three regimes 
without the pump field, the Berry curvatures of the conduction and valence bands are $\mp \eta \Delta \text{v}_0^2/\{4[(\text{v}_0p)^2+(\Delta/2)^2]^{3/2}\}$. In the presence of the pump field, the signs of the Berry curvatures of the renormalized bands should remain the same. Therefore, the Hall conductivity contribution from the conduction (valence) electrons will be negative (positive) at valley $K$ and positive (negative) at valley $K^{\prime}$. The negative sign of $\sigma_{xy}$ then follows from the larger population of excited carriers at valley $K$ in comparison with valley $K^{\prime}$ due to the valley selection rule. 

Using Eq.~\eqref{eq8} and taking the sum over $\eta$-valley dependent contributions, we can numerically calculate the total Hall conductivity as a function of pump frequency $\omega$ (see solid curves in Fig.~\ref{Fig3}). We notice a 
%With the assumption of a momentum-independent value of $\delta_{\eta}$ the complete dependencies of $\sigma_{xy}(\omega)$ for the $K$ valley ($\eta=1$) were  numerically calculated by using  Eq.~(\ref{eqMain}). The results for three different non-equilibrium regimes are presented in Fig.~\ref{ana_num}. 
similar behavior in regimes I and III, namely, an abrupt increase in the absolute value of conductivity as the frequency approaches $\Delta/\hbar$ and a further smooth decrease of $\sigma_{xy}$.
%excompares the above analytic results for the valley $K$ with numerical results obtained from evaluating Eq.~(\ref{eqMain}) using . We see that there is an excellent agreement for frequency values $\omega > \Delta$ between the analytic and numerical results for all three regimes, with the two sets of results completely overlapping with each other. 
%As shown, regimes I and III behave very similarly; their corresponding results for the valley $K^{\prime}$ are also similar to each other, having $\sigma_{\eta = -1,xy} \approx -e^2/2h$  because the valley $K^{\prime}$ is approximately uncoupled to the pump field with $\delta_{\eta = -1} \approx 0$. Summing the contributions from both valleys therefore yield a total Hall conductivity for regimes I and III with a similar profile as in Fig.~\ref{ana_num}, except shifted by $-0.5e^2/h$. 

The most significant feature is that regime II shows a completely different behavior (inset of Fig.~\ref{Fig3}) by revealing a dramatically enhanced Hall conductivity. Indeed, the ratio $\sigma^{\textrm{(II)}}/\sigma^{\textrm{(I)}}$ for similar parameters is on the order of $10^5$. Further, it saturates at large frequencies to a value independent of applied EMF power. This is a direct consequence of the inversion of electron population in regime II.

Next, we take into account the full momentum dependence of the light-matter coupling at both valleys $K$ and $K^{\prime}$ by using the exact relations, Eqs.~(\ref{eq6}) and (\ref{eqSuper}), for a left circularly polarized pump field ($\sigma=1$). Now, the light couples strongly to the $K$ valley and weakly to the $K^\prime$ valley, inducing an enhanced dynamical gap at the $K$ valley with $\delta_1 > \delta_{-1}$ (see dashed lines in Fig.~\ref{Fig3}). A crucial assumption of these calculations is that \textit{both the valleys are described by the same type of steady state distribution functions}, regardless of different values of their dynamical gaps.   

Accounting for the small dynamical gap in the $K^\prime$ valley leads to minute changes of $\sigma_{xy}(\omega)$ in regimes I and III (compare the dashed and solid red (blue) curves in Fig.~\ref{Fig3}). However, the results obtained for regime II are drastically different. They show a small, sharp peak when $\hbar \omega=\Delta $ (compare the dashed green curve in main plot and solid green curve in Inset of Fig.~\ref{Fig3}). We explain this behavior as a consequence of the crucial assumption that nonequilibrium distributions are realized in both valleys. Observation of the frequency dependence of $\sigma_{xy}$ enables us to distinguish between the different nonequilibrium steady states under an optical pump field.
%
%present case where the intraband relaxation is much faster than the interband relaxation both valleys are (\textit{i.e.}, regime II).

%---------------------------------
%---------------------------------
%---------------------------------

\subsection{Spin-orbit coupling effects in TMDs}
Finally, using the full $k\cdot p$ Hamiltonian~\eqref{eq1} of TMDs, we include spin-orbit coupling effects in our analysis.
%Total Hall conductivity.---}
%Numerical results of the total Hall conductivity  ---}
Typical parameters of MoS$_2$ monolayer \cite{xiao2012coupled} are employed: $\Delta=1.66\,\textrm{eV}$ and $\lambda_\textrm{so}=75\,\textrm{meV}$.
%Our calculations here The left circularly polarized light couples strongly to the $K$ valley and weakly to the $K^\prime$ valley, inducing an enhanced dynamical gap at the $K$ valley with $\delta_1 > \delta_{-1}$ . 
%In our theory $\tau$ and $\tau_{\mathrm{rec}}$ are phenomenological parameters and taken to have the same values at the two valleys.
Calculation results for the three regimes are presented in Fig.~\ref{Fig4}a--c. As expected, SOI results in the appearance of a second threshold in the conductivity of regimes I and III (Fig.~\ref{Fig4}a and Fig~\ref{Fig4}c) and two sharp peaks in regime II (Fig.~\ref{Fig4}b), once the EF frequency reaches the band gap values $\Delta \pm \lambda_{so}$ (at $\hbar \omega=1.585 eV$ and $\hbar \omega=1.735 eV$) for the two spin-split bands \cite{Remark_excitons}. The plots also demonstrate the dependence of $\sigma_{xy}$ on the value of the gap, $\delta_{\eta}(0)$. It is important to note, that with account of the SOI, there opens a possibility to established spin-polarized Hall conductivity if $\omega$ is in the narrow frequency interval $(\Delta-\lambda_\textrm{so},\Delta+\lambda_\textrm{so})$. Indeed, at the first threshold (see Figs.~\ref{Fig4}) due to the energy conservation, there will be Hall current of electrons and holes with a predefined projection of spin~\cite{Glaz}.

\section{Conclusions} 

We have developed a theory for the photovoltaic valley-dependent Hall effect in a two-dimensional Dirac semiconductor driven by a strong electromagnetic field. We have found that the valley-dependent Hall conductivity is strongly enhanced when the pump field frequency is close to the transition energies of the two spin-split bands at $K$ and $K^{\prime}$ valleys.  
We have also shown that the conductivity is highly sensitive to nonequilibrium carrier distribution functions due to the joint influence of the pump field and the intraband relaxation and interband recombination processes. 

\section*{Acknowledgements}

We thank Yuri Rubo and Sergej Flach for useful discussions, Joel Rasmussen (RECON) for a critical reading of our manuscript, and Ekaterina Savenko for help with the figures.
V.~M.~K. has been supported by the Russian Science Foundation (Project No. 17-12-01039).
W.~K.~T. acknowledges the support from startup funds of the University of Alabama. 
M.~V.~F. has been partially supported by the Ministry of Education and Science of the Russian Federation in the framework of the Increase Competitiveness Program of NUST MISIS K2-067-2018. 
I.~G.~S. and M.~V.~F. acknowledge the support of the Institute for Basic Science in Korea (Project No.~IBS-R024-D1).

%and  and the Russian Science Foundation (Contract $No. 16-12-00095$).

%\newpage
%\bibliography{NLHE}

\begin{thebibliography}{50}
\bibitem{Remark_excitons} Excitonic effects are disregarded in our model. Inclusion of excitonic effects is expected to bring additional resonant features in the photovoltaic Hall conductivity at the exciton energies associated with the two spin-split bands.
\end{thebibliography}
%\bibliographystyle{apsrmp4-1}

%---------
%---------
%---------

%---------
%---------
%---------

%---------
%---------
%---------

%
\begin{widetext}

\appendix

\section{RWA, the dispersion of PDQs and the Hall conductivity via the Keldysh approach}
\label{ApA}

Here we present a detailed discussion of the rotating wave approximation (RWA) used; properties of the gapped dispersion of PDQs; calculation of the general expression for the conductivity using a nonequilibrium Keldysh approach. 

Assuming that the probe field is weak, the current can be calculated as a linear response to this field:
\begin{gather}\label{eqsp3}
j_\alpha(t)=\int_{C} dt'Q_{\alpha\beta}(t,t')\mathcal{A}_{\beta}(t'), 
~~\textrm{where}~~
Q_{\alpha\beta}(t,t_1)=-i\frac{e^2}{c}\,\textmd{Tr}\left[\hat{\textrm{v}}_\alpha G(t,t_1)\hat{\textrm{v}}_\beta G(t_1,t)\right]_{C},
\end{gather}
where $C$ is the Keldysh contour. The Green's functions in Eq.~(\ref{eqsp3}) should be calculated in the presence of the pump field accounting for it in unperturbed manner. Thus, in contrast to standard linear response technique, Green's functions in~(\ref{eqsp3}) are principally nonequilibrium and in general case, they depend on both the times $t$ and $t'$ separately. Thus, the Green's function satisfies the following equation:
\begin{gather}\label{eqsp4}
\left[i\partial_t-H_0-H_{\textrm{int}}(t)\right]G(t,t')=\delta_{t,t'}.
\end{gather}
It is written in a pseudospin representation of the  operators $\hat{\sigma}_\alpha$. This representation is not very convenient. Therefore using a unitary transformation, we switch to another representation using the conduction and valence bands:
\begin{gather}\label{eqsp5}
\tilde{H}_0=U_p^\dag H_0U_p=\left(
                              \begin{array}{cc}
                                \varepsilon_c(p) & 0 \\
                                0 & \varepsilon_v(p) \\
                              \end{array}
                            \right),~~\textrm{where}\\
                            \nonumber
\varepsilon_{c,v}(p)=\frac{s\eta\lambda_\textrm{so}}{2}\pm|E_{s\eta}(p)|,\,\,\,\, E_{s \eta }(p) =\pm \sqrt{(\text{v}_0 p)^2+
 \left (\frac{\Delta-s\eta \lambda_\textrm{so}}{2} \right )^2},\\\nonumber
U_p = \begin{bmatrix} \cos(\theta/2) & \sin(\theta/2) \\
 \sin(\theta/2)e^{i\eta\phi} &
 -\cos(\theta/2)e^{i\eta\phi}
 \end{bmatrix},\,\,\,\,
 \nonumber
 U^\dag_p = \begin{bmatrix} \cos(\theta/2) & \sin(\theta/2)e^{-i\eta\phi} \\
 \sin(\theta/2) &
 -\cos(\theta/2)e^{-i\eta\phi}
 \end{bmatrix},\\
 \nonumber
 U^\dag_pU_p=U_pU^\dag_p=1,
\end{gather}
where $\theta$ is the polar angle with
\begin{eqnarray}\label{EqCosAp}
\cos\theta &=& \frac{(\Delta-s\eta\lambda_\textrm{so})/2}{|E_{s\eta}(p)|},
\nonumber \\
\sin\theta &=& \frac{\eta \text{v}_0 p}{|E_{s\eta}(p)|},
\end{eqnarray}
and $\phi = \tan^{-1}(p_y/p_x)$. Applying this transformation to Eq.~(\ref{eqsp4}), we find
\begin{gather}\label{eq9}
\left[i\partial_t-\tilde{H}_0-\tilde{V}(t)\right]\tilde{G}(t,t')=\delta_{t,t'},~\textrm{where}~~
\tilde{G}(t,t')=U_p^\dag G(t,t')U_p,\\\nonumber
\tilde{V}(t)=U_p^\dag H_{\textrm{int}}(t)U_p=\left(
                                            \begin{array}{cc}
                                              \tilde{V}_{cc}(t) & \tilde{V}_{cv}(t) \\
                                              \tilde{V}_{vc}(t) & \tilde{V}_{vv}(t) \\
                                            \end{array}
                                          \right).
\end{gather}
\begin{figure}[!t]
\includegraphics[width=7cm, height=7cm]{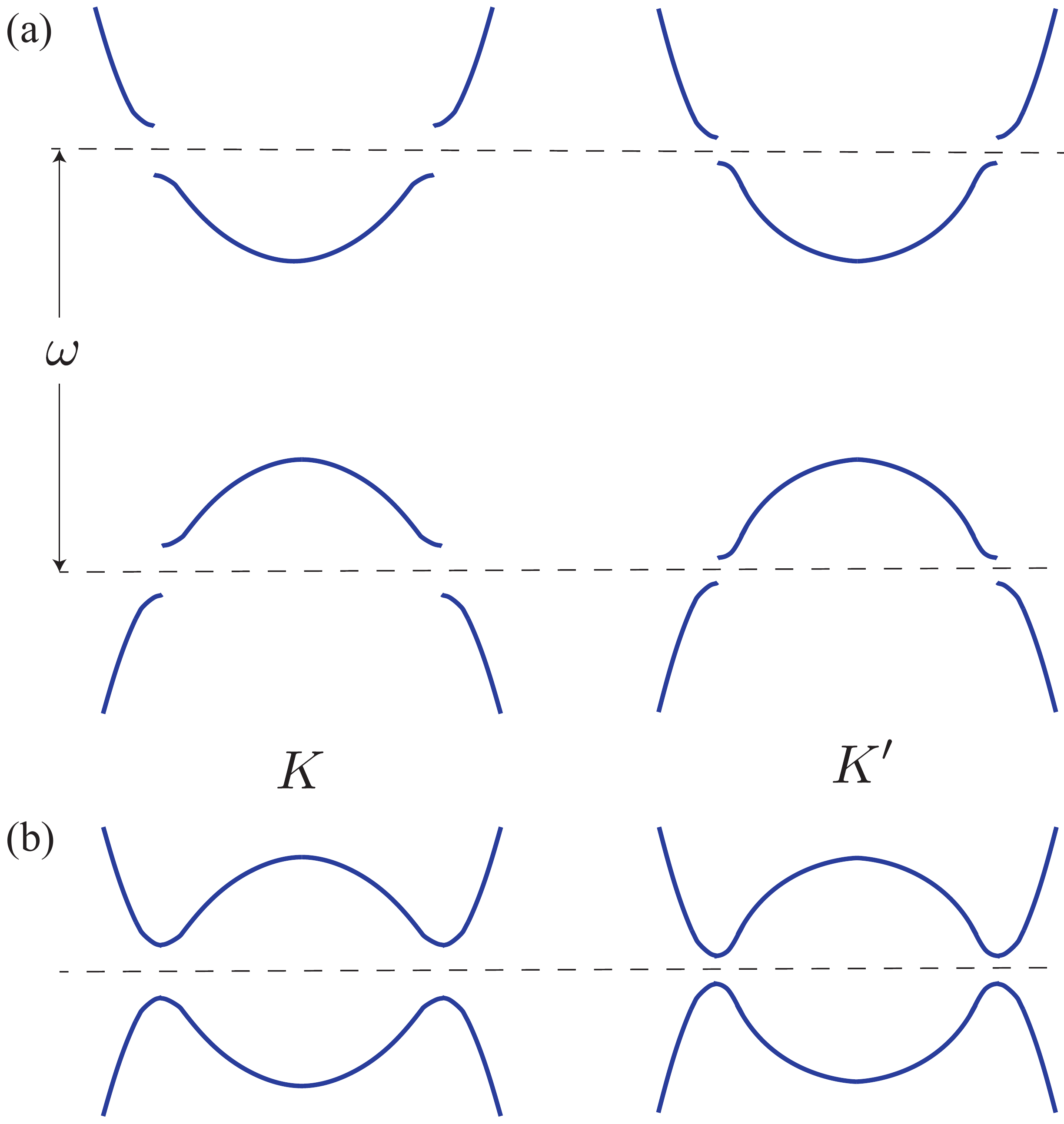}
%[width=7cm, height=5cm]{Fig2.pdf}
\caption{(a) Dynamical gaps in the rotating-frame conduction and valence bands under a left circularly polarized light. For clarity, here we only show the pair of energy dispersions for spin up electrons at valley $K$ and spin down electrons at valley $K^{\prime}$, which have a band gap $\Delta-\lambda_{\mathrm{so}}$. The other pair of energy dispersions with a band gap $\Delta+\lambda_{\mathrm{so}}$ will have dynamical gaps at a different value of momentum. (b) The corresponding quasienergy spectra  Eq.~(\ref{eq5}).}
\label{DynGap}
\end{figure}
The pump field reads $\textbf{A}(t)=\textbf{A}e^{-i\omega t}+\textbf{A}^*e^{i\omega t}$.
After the transformation into cv-basis, we find
\begin{gather}\label{eq11A}
\tilde{V}(t)
=\left(
                                            \begin{array}{cc}
                                              \tilde{V}_{cc}(t) & \tilde{V}_{cv}(t) \\
                                              \tilde{V}_{vc}(t) & \tilde{V}_{vv}(t) \\
                                            \end{array}
                                          \right)=\frac{e}{c}\textbf{A}\cdot\left(
                       \begin{array}{cc}
                         \tilde{\textbf{v}}_{cc} & \underline{\tilde{\textbf{v}}_{cv}} \\
                         \tilde{\textbf{v}}_{vc} & \tilde{\textbf{v}}_{vv} \\
                       \end{array}
                     \right)e^{-i\omega t}+\frac{e}{c}\textbf{A}^*\cdot\left(
                       \begin{array}{cc}
                         \tilde{\textbf{v}}_{cc} & \tilde{\textbf{v}}_{cv} \\
                        \underline{ \tilde{\textbf{v}}_{vc}} & \tilde{\textbf{v}}_{vv} \\
                       \end{array}
                     \right)e^{i\omega t}.
\end{gather}
Here the diagonal (intraband cc and vv) terms 
result in a nonresonant change of the spectrum of quasienergies, leading to small corrections to the Green's functions proportional to $\frac{e^2}{c^2}\tilde{\text{v}}_{ii}^2A^2/(\textrm{min}\{\Delta,\hbar\omega\})^2$, where $i=$c or v and we assume $|\frac{e}{c}\tilde{\text{v}}_{ii}A|\ll\textrm{min}\{\Delta,\hbar\omega\}$. Consequently they can be disregarded. 
The off-diagonal (interband cv, vc) terms have both the resonant and nonresonant contributions (within the RWA approach) and for similar reasons, we will keep only the resonant ones underlined in Eq.~\eqref{eq11A}. 

The resulting equation for the Green's function reads
\begin{gather}\label{eq12}
\left(
  \begin{array}{cc}
    i\partial_t-\varepsilon_c(\textbf{p}) & -\frac{e}{c}\tilde{\textbf{v}}_{cv}\cdot\textbf{A}e^{-i\omega t} \\
    -\frac{e}{c}\tilde{\textbf{v}}_{vc}\cdot\textbf{A}^*e^{i\omega t} & i\partial_t-\varepsilon_v(\textbf{p}) \\
  \end{array}
\right)\tilde{G}(t,t')=\delta_{t,t'},\\\nonumber
\tilde{\textbf{v}}_{cv}\cdot\textbf{A}=\text{v}_0[\sin^2(\theta/2)e^{-i\eta\phi}(\eta A_x+iA_y)
+\cos^2(\theta/2)e^{i\eta\phi}(-\eta A_x+iA_y)],\\\nonumber
\tilde{\textbf{v}}_{vc}\cdot\textbf{A}^*=\text{v}_0[\sin^2(\theta/2)e^{i\eta\phi}(\eta A^*_x-iA^*_y)
+\cos^2(\theta/2)e^{-i\eta\phi}(-\eta A^*_x-iA^*_y)].
\end{gather}
Important to note, that $\tilde{\textbf{v}}_{cv}$ and $\tilde{\textbf{v}}_{vc}$ depend on the momentum, $\textbf{p}$.
In order to transform into a rotating reference frame, we utilize the operator
\begin{eqnarray}\label{eq13}
S(t)=\left(
       \begin{array}{cc}
         e^{-i\omega t/2} & 0 \\
         0 & e^{i\omega t/2} \\
       \end{array}
     \right),
\end{eqnarray}
and using $g(t-t')=S^\dag(t)\tilde{G}(t,t')S(t')$, we find
\begin{eqnarray}\label{eq14}
\left(
  \begin{array}{cc}
    i\partial_t-\varepsilon_c(\textbf{p})+\hbar\omega/2 & -\frac{e}{c}\tilde{\textbf{v}}_{cv}\cdot\textbf{A} \\
    -\frac{e}{c}\tilde{\textbf{v}}_{vc}\cdot\textbf{A}^* & i\partial_t-\varepsilon_v(\textbf{p})-{\hbar\omega}/2 \\
  \end{array}
\right)g(t-t')=\delta_{t,t'}.
\end{eqnarray}
Note, this equation is translational invariant in time.
%%%%%
%%%%%%
%%%%%

The resulting quasienergy spectrum is given by
\begin{gather}\label{eq5A}
\varepsilon_{1,2}(\mathbf{p})=\pm\sqrt{\left(|E_{s \eta }(p)|-\frac{\hbar \omega}{2}\right)^2+\vert \delta_\eta(\mathbf{p})\vert^2}.
\end{gather}
As indicated in Fig.~\ref{DynGap}, dynamical gaps open in each of the four copies of gapped Dirac dispersions in the TMD band structure; the valley selection rule causes the dynamical gap at one valley to dominate. The magnitude of the dynamical gap then captures the seminal valley-dependent selection rule~\cite{xiao2012coupled} and becomes isotropic in the momentum space, $|\delta_\eta({p})|= (e\text{v}_0/c)(1+\eta m \cos\theta_p)$, for circularly-polarized pump field $\textbf{A}(t)$.

Let us transform Eq.~(\ref{eqsp3}) into the band representation, leaving only interband matrix elements of the operators $\hat{\text{v}}_\alpha,\,\hat{\text{v}}_\beta$. Calculating the trace $\textmd{Tr}[...]$, we find the sum of terms containing diagonal matrix elements like $G_{cc}(t,t')G_{vv}(t',t)$ and non-diagonal elements like $G_{cv}(t,t')G_{vc}(t',t)$. Indeed, the calculation of trace results in the expression:
\begin{gather}\label{Add1}
Q_{xy}(t,t')=-i\frac{e^2}{c}\textmd{Tr}\left[\tilde{\text{v}}_x\tilde{G}(t,t')\tilde{\text{v}}_y\tilde{G}(t',t)\right]
=Q^{(1)}_{xy}(t,t')+Q^{(2)}_{xy}(t,t'),
\end{gather}
where
\begin{gather}\label{Add2}
Q^{(1)}_{xy}(t,t')=-i\frac{e^2}{c}\left[\tilde{v}^x_{cv}\tilde{G}_{vv}(t,t')\tilde{v}^y_{vc}\tilde{G}_{cc}(t',t)+
\tilde{v}^x_{vc}\tilde{G}_{cc}(t,t')\tilde{v}^y_{cv}\tilde{G}_{vv}(t',t)\right],\\\nonumber
Q^{(2)}_{xy}(t,t')=-i\frac{e^2}{c}\left[\tilde{v}^x_{cv}\tilde{G}_{vc}(t,t')\tilde{v}^y_{cv}\tilde{G}_{vc}(t',t)+
\tilde{v}^x_{vc}\tilde{G}_{cv}(t,t')\tilde{v}^y_{vc}\tilde{G}_{cv}(t',t)\right],~\textrm{where}\\
\nonumber
\tilde{v}^x_{cv}=\text{v}_0\eta(-\cos^2(\theta/2)e^{i\eta\varphi_{\textbf{p}}}+\sin^2(\theta/2)e^{-i\eta\varphi_{\textbf{p}}}),\\
\nonumber
\tilde{v}^x_{vc}=\text{v}_0\eta(-\cos^2(\theta/2)e^{-i\eta\varphi_{\textbf{p}}}+\sin^2(\theta/2)e^{i\eta\varphi_{\textbf{p}}}),\\\nonumber
\tilde{v}^y_{cv}=i\text{v}_0(\cos^2(\theta/2)e^{i\eta\varphi_{\textbf{p}}}+\sin^2(\theta/2)e^{-i\eta\varphi_{\textbf{p}}}),\\\nonumber
\tilde{v}^y_{vc}=i\text{v}_0(-\cos^2(\theta/2)e^{-i\eta\varphi_{\textbf{p}}}-\sin^2(\theta/2)e^{i\eta\varphi_{\textbf{p}}}),
\end{gather}
and Green's function in band representation accounting for the strong pump field reads
\begin{gather}\label{eqsp27}
\tilde{G}(t,t')=S(t)g(t-t')S^\dag(t')=\left(
                          \begin{array}{cc}
                            g_{cc}(t-t')e^{-i\frac{\omega}{2}(t-t')} & g_{cv}(t-t')e^{-i\frac{\omega}{2}(t+t')} \\
                            g_{vc}(t-t')e^{i\frac{\omega}{2}(t+t')} & g_{vv}(t-t')e^{i\frac{\omega}{2}(t-t')} \\
                          \end{array}
                        \right).
\end{gather}
Substituting Eq.~\eqref{eqsp27} in~\eqref{Add2} and performing the Fourier transform in time, we find that non-diagonal terms of $Q^{(2)}_{xy}(t,t')$ are proportional to $e^{\pm 2i\omega t}$, whereas the diagonal ones of $Q^{(1)}_{xy}(t,t')$ do not contain frequency-dependent exponents. 
Thus $Q^{(2)}_{xy}(t,t')$ term (describing the second harmonic generation effects) should be further disregarded in the framework of the RWA we use. 
%(since these terms give zero current after averaging over time). 

The probe field depends on time as $\mbox{\boldmath{$\mathcal{A}$}}(t)=\mbox{\boldmath{$\mathcal{A}$}}\cos(\Omega t)$, producing the in-plane current
\begin{gather}\label{Add3}
j_x(t)=Q_{xy}(\omega,\Omega)\mathcal{A}_y(\Omega)e^{-i\Omega t},\\\nonumber
Q_{xy}(\omega,\Omega)=\eta\sum_{\textbf{p},\varepsilon}\left(F^*(\textbf{p})
\left[g_{vv}(\textbf{p},\varepsilon+\omega+\Omega)g_{cc}(\textbf{p},\varepsilon)\right]^<-F(\textbf{p})
\left[g_{cc}(\textbf{p},\varepsilon-\omega+\Omega)g_{vv}(\textbf{p},\varepsilon)\right]^<\right),\\\nonumber
F(\textbf{p})=\cos\theta-i\frac{\eta}{2}\sin^2\theta\sin(2\phi).
\end{gather}
All the Green's functions here depend on the absolute value of particle momentum $\textbf{p}$ thus, the term $\sin(2\phi)$ in $F(\textbf{p})$ does not play the role due to the angle integration. The structure of $[gg]^<=g^Rg^<+g^<g^A$ contains the lesser $g^<$ and retarded/advansed $g^{R,A}$ functions which can be easily found from expression \eqref{eq14}.

%Taking the limit $\Omega\rightarrow 0$ and  averaging over time, we conclude that only the expression containing diagonal elements of the Green's functions give nonzero result.

The time-averaged Hall current is expressed via the Hall conductivity $\sigma_{xy}$ as $j_x=\sigma_{xy}\mathcal{E}_y$, {and $\mathcal{E}_y(t)=-\frac{1}{c}\partial_t\mathcal{A}_y(t)$ } {is the probe electric field} taken to be along the $y$ axis. The Hall conductivity $\sigma_{xy}$ contains nonlinear effects due to the presence of a strong pump EMF. Taking the integration over $\varepsilon$ in~\eqref{Add3}, we find (in the limit of a static probe field $\Omega \rightarrow 0$) a generic expression for the \textit{photovoltaic valley-dependent Hall conductivity}~\eqref{eqSuper}.
%
%\begin{gather}\label{eqSuper}
%\sigma_{xy} = \frac{2e^2\text{v}_0^2}%\left[\frac{u^4_{{p}}}{(\varepsilon_1({p})-\varepsilon_2({p})+\hbar\omega)^2}-\frac{v^4_{{p}}}
%{(\varepsilon_1({p})-\varepsilon_2({p})-\hbar\omega)^2}\right],
%\end{gather}
%

%--------------------------------
%--------------------------------
%--------------------------------

\section{Valley-resolved Hall conductivity}
\label{ApB}

Here we present a detailed discussion of  conductivity in the regime where analytical treatment is possible and a single-valley contribution results.
In order to focus on the essential valley-resolved physics (and obtain analytic results for the photovoltaic valley-dependent Hall conductivity), we will disregard spins and spin-orbit interaction and the  dependence of  $\delta_\eta$ on momentum $p$ in present section.
With these approximations the Hall conductivity at valley $\eta$ can be expressed as in Eq.~\eqref{eq8}.

In the limit of vanishing pump EMF, \textit{i.e.} $\delta_\eta \to 0$, the distribution functions of conduction and valence band electrons reduce to those in equilibrium, we have $f_c({p})=0$ and $1-2n_1({p})=1$
so that Eq.~(\ref{eq8})  recovers the correct value of the \textit{DC} Hall conductivity of a single valley $\sigma_{xy}= \eta e^2/(4\pi \hbar)$ for 2D Dirac semiconductor~\cite{Asgari}.

Substituting the expressions for $n_{1,2}(p)$ corresponding to the three regimes, Eq.~(\ref{eq8}) can be written as
%For these three regimes the $\eta$-valley contribution to the  Hall conductivity is written as
%contribution Taking the sum over the two valleys, we find the following total valley-dependent Hall conductivity in the three regimes
%
\begin{gather}
\label{eqMainAB}
\left(
                   \begin{array}{c}
                   \sigma_{\eta,~ xy}^{(\mathrm{I})}\\
                    \sigma_{\eta,~ xy}^{(\mathrm{II})} \\
                    \sigma_{\eta,~ xy}^{(\mathrm{III})} \\
                   \end{array}
                 \right)
=\eta \frac{e^2{\Delta}}{32\pi \hbar}\int\limits_{-(\omega-{\Delta})/2}^\infty
d\xi
\left(
  \begin{array}{c}
  \textmd{sign}\xi\\
    1 \\
    \frac{\xi}{\sqrt{\xi^2+\delta_\eta^2}} \\
  \end{array}
\right)\mathcal{F}_\omega(\delta_\eta,\xi).
\end{gather}
%
% %
% \begin{gather}
% \label{eqMain}
% \left(
%                    \begin{array}{c}
%                    \sigma_{\eta,~ xy}^{(\mathrm{I})}\\
%                     \sigma_{\eta,~ xy}^{(\mathrm{II})} \\
%                     \sigma_{\eta,~ xy}^{(\mathrm{III})} \\
%                    \end{array}
%                  \right)
% =\eta \frac{e^2{\Delta}}{32\pi \hbar}\int\limits_{-(\omega-{\Delta})/2}^\infty
% d\xi\\\nonumber
% \times
% \left[\left(
%   \begin{array}{c}
%   \textmd{sign}\xi\\
%     1 \\
%     \frac{\xi}{\sqrt{\xi^2+\delta_\eta^2}} \\
%   \end{array}
% \right)\mathcal{F}_\omega(\delta_\eta,\xi)-(\textmd{sign}\xi)\mathcal{F}_\omega(0,\xi)\right]+\frac{\eta e^2}{4\pi \hbar}.
% \end{gather}
% %
Further analytic progress can be made if we focus on the frequency range $\vert \hbar\omega-\Delta \vert \gg \delta_\eta$. If the frequency of the pump EMF is below the gap, ${\Delta}-\hbar\omega \gg \delta_\eta$, only virtual transitions between the conduction and valence bands
%in the K valley
occur, resulting in a renormalization of the band energies (\textit{i.e.}, the dynamic Stark effect) as described by the quasienergies  $\varepsilon_{1,2}(p)$ of the PDQs. This scenario is described by regime I.
\begin{figure}[!t]
\includegraphics[width=9cm, height=6cm]{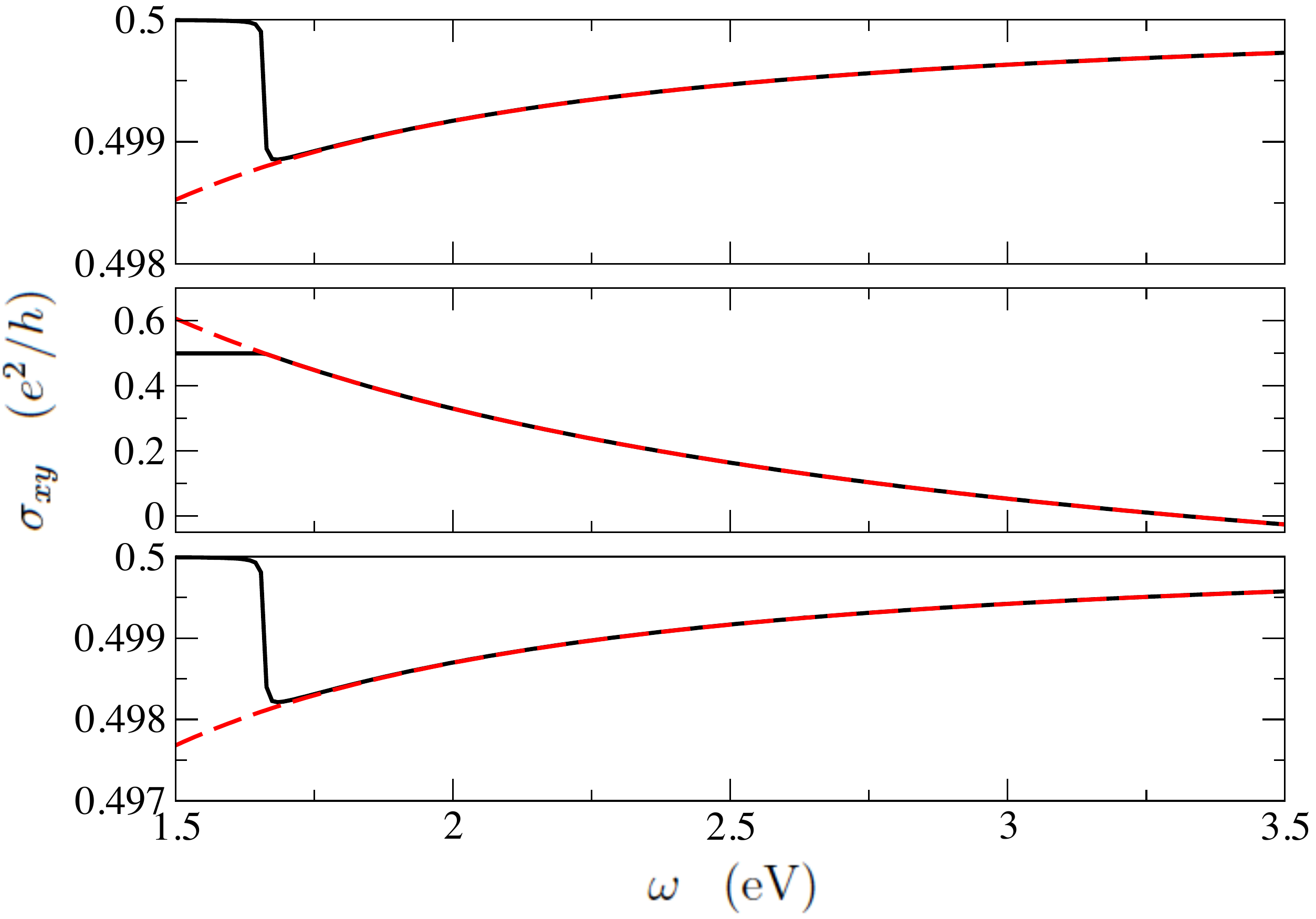}
%[width=7cm, height=4cm]{Fig5.pdf}
\caption{Single-valley contribution to the Hall conductivity $\sigma_{\eta = 1,xy}(\omega)$ from the valley $K$ in regimes I, II, III (upper, middle, and lower panels, respectively). Numerical results using Eq.~(\ref{eqMain}) with a constant $\delta_{\eta = 1}=1\,\mathrm{meV}$ and and $\Delta =1.66\,\mathrm{eV}$ are indicated by black solid lines and the corresponding analytic results using Eqs.~(\ref{eq13_1})-(\ref{eq13_3}) are shown by red dashed lines. The analytic result Eq.~(\ref{eq13_2}) for regime II also contains a sharp peak at $\hbar\omega = \Delta$, which is not shown as it is beyond the resolution of the plot.
%The parameters used are $\delta_{\eta = 1}=1\,\mathrm{meV}$ and $\Delta =1.66\,\mathrm{eV}$.
} \label{ana_num}
\end{figure}
%
%
%
%and it results just in the renormalization of  PDQs energies $\epsilon_{1,2}(p)$,  i.e. the  \textit{dynamic Stark effect}. However, in this frequency range the regime I  can be %realized only.
Calculating the integral over $\xi$ in Eq.~(\ref{eqMainAB}), we obtain {the $\eta$-valley contribution to the Hall conductivity} as
\begin{gather}\label{eq11B}
\sigma_{\eta,~xy}^{(\mathrm{I})}=\frac{\eta e^2}{4\pi \hbar}\left[1-\frac{2\delta^2_{\eta}}{{\Delta}({\Delta}-\hbar\omega)}\right].
\end{gather}
%
%
% \begin{gather}\label{eq11}
% \sigma_{\eta,~xy}^{(\mathrm{I})}=-\eta\frac{e^2}{2\pi \hbar}\frac{\delta^2_R}{{\Delta}({\Delta}-\omega)}+\frac{\eta e^2}{4\pi \hbar}.
% %\sigma_{xy}^{(\mathrm{I})}=-\frac{e^2}{4\pi}\frac{2\delta^2_R}{{\Delta}({\Delta}-\omega)},~~{\Delta}-\omega \gg 2\delta_R
% \end{gather}
%
In the opposite limit when the frequency $\omega$ exceeds the gap $\Delta$, interband transitions occur and all the three regimes can be established.
\begin{figure}[!b]
\includegraphics[width=9cm, height=6cm]{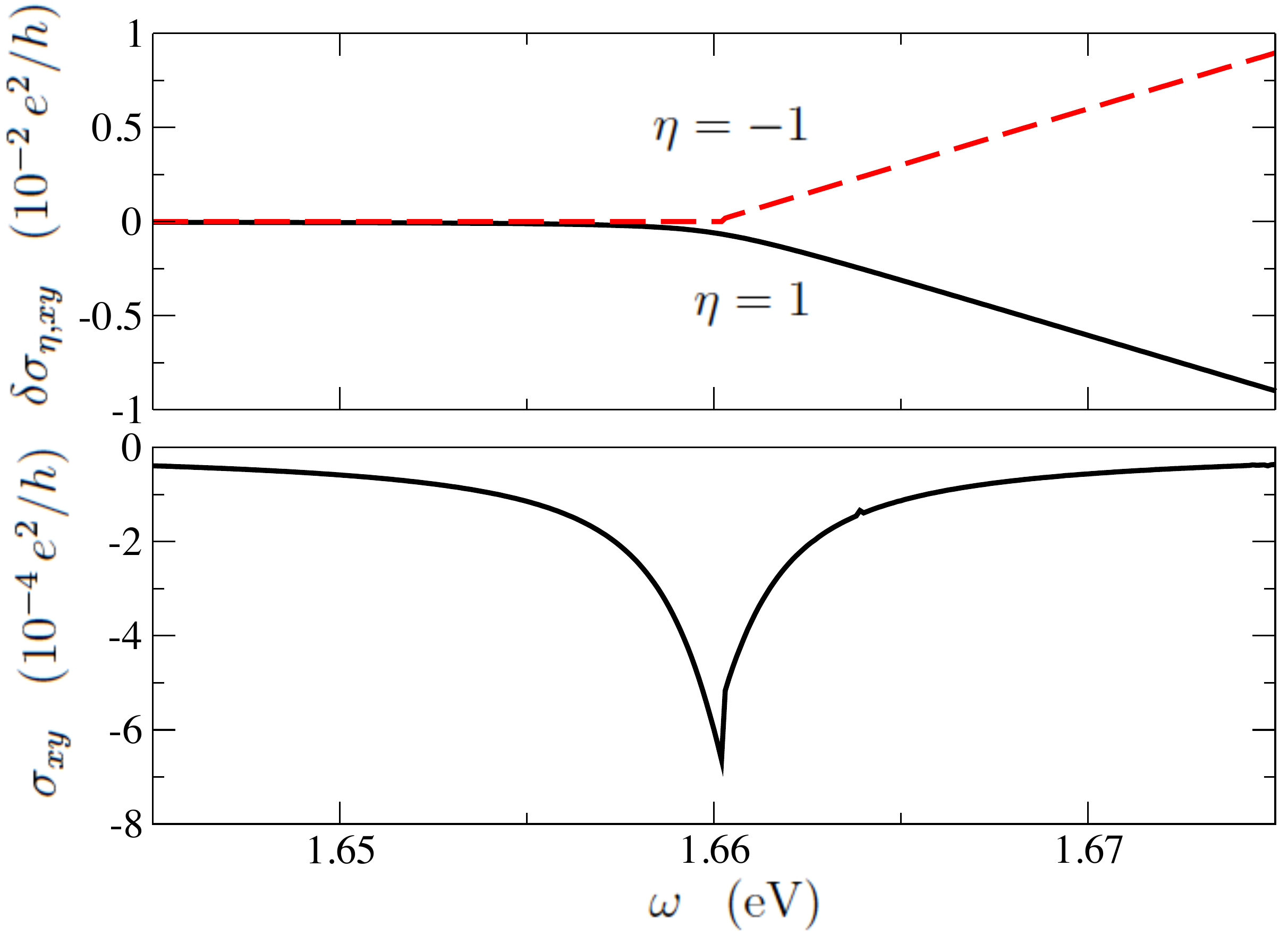}
\caption{Hall conductivity in regime II calculated from Eq.~(\ref{eqMain}) including the full momentum dependence of $\delta_{\eta}(p)$. Upper panel: single-valley Hall conductivity contribution rendered from the respective values without pump field at valleys $K$ and $K^{\prime}$, $\delta\sigma_{\eta,xy}(\omega) = \sigma_{\eta,xy}(\omega)-\eta e^2/2h$. Lower panel: the sum of the two graphs yield the total Hall conductivity $\sigma_{xy} =  \delta\sigma_{\eta = 1,xy}+\delta\sigma_{\eta = -1,xy}$. Parameters used are the same as in Fig.~\ref{ana_num}.
} \label{peak_regime2}
\end{figure}

Calculating the integral over $\xi$ in Eq.~(\ref{eqMainAB}), we arrive at the following results:
\begin{gather}
\sigma^{(\mathrm{I})}_{\eta,~xy}(\omega)= \frac{\eta e^2}{4\pi \hbar}\left(1-\frac{4\Delta \delta_\eta}{(\hbar\omega)^2}\right) ,\label{eq13_1} \\
%-\frac{\eta e^2}{\pi \hbar} \frac{\Delta \delta_\eta}{\omega^2}+\frac{\eta e^2}{4\pi \hbar} ,\label{eq13_1} \\
\sigma^{(\mathrm{II})}_{\eta, ~xy}(\omega)=\frac{\eta e^2}{4\pi \hbar}\left[-1+\frac{{2\Delta}}{\hbar\omega}-\frac{2\delta_\eta^2}{\Delta(\hbar\omega-\Delta)} \right],\\
%\frac{\eta e^2}{2\pi \hbar}\left[\frac{{\Delta}}{\omega}-\frac{\delta_\eta^2}{\Delta(\omega-\Delta)} \right]\label{eq13_2},\\
\sigma^{(\mathrm{III})}_{\eta,~xy}(\omega)=\frac{\eta e^2}{4\pi \hbar}\left(1-\frac{2\pi \Delta\delta_\eta}{(\hbar\omega)^2}\right). 
\end{gather}

Figure~\ref{ana_num} compares the above analytic results for the valley $K$ with numerical results obtained from evaluating Eq.~(\ref{eqMainAB}) using a momentum-independent value of $\delta_{\eta}$. We see that there is an excellent agreement for frequency values $\omega > \Delta$ between the analytic and numerical results for all the three regimes, with the two sets of results completely overlapping with each other. As shown, regimes I and III behave very similarly. Their corresponding results for the valley $K^{\prime}$ are also similar to each other, having $\sigma_{\eta = -1,xy} \approx -e^2/2h$ since the valley $K^{\prime}$ is approximately uncoupled to the pump field with $\delta_{\eta = -1} \approx 0$. Summing up the contributions from both valleys yields the total Hall conductivity for regimes I and III with a similar profile as in Fig.~\ref{ana_num}, except shifted by $-0.5e^2/h$ (see main text, Fig.~3).

Regime II shows a completely different behavior since both valleys are in the saturated state with a large inverted population of conduction band electrons. The corresponding numerical result for the valley $K^{\prime}$ is approximately equal in magnitude and opposite in sign to that for the valley $K$ in Fig.~\ref{ana_num}, except for a near-resonance region $\omega \approx \Delta$. In the vicinity of resonance (see Fig.~\ref{peak_regime2}) we find that the Hall conductivity at the valley $K$ exceeds in magnitude that at the valley $K^{\prime}$, resulting in a very sharp peak at $\hbar\omega = \Delta$.

Under a left circularly polarized pump field ($\sigma = 1$), the total Hall conductivity is negative in all the three regimes.
This can be understood from the renormalized band structures in the rotating frame (Fig.~\ref{DynGap}). Without the pump field, the Berry curvatures of the conduction and valence bands are $\mp \eta \Delta \text{v}_0^2/\{4[(\text{v}_0p)^2+(\Delta/2)^2]^{3/2}\}$. In the presence of the pump field, the signs of the Berry curvatures of the renormalized bands should remain the same, and the Hall conductivity contribution due to the conduction (valence) band will be negative (positive) at the valley $K$ and positive (negative) at the valley $K^{\prime}$. The negative sign of $\sigma_{xy}$ follows due to larger population of excited carriers at valley $K$ than at valley $K^{\prime}$ coming from the valley selection rule.

\end{widetext}

\end{document}